# Entangled Two-Photon Absorption by Atoms and Molecules:
# A Quantum Optics Tutorial

*Michael G. Raymer,[1,2]\* Tiemo Landes,[1,2] Andrew H. Marcus [2,3]*

[1.] Department of Physics, University of Oregon, Eugene, OR 97403, USA

[2.] Oregon Center for Optical, Molecular and Quantum Science, University of Oregon, Eugene, OR 97403, USA

[3.] Dept of Chemistry and Biochemistry, University of Oregon, Eugene, OR 97403, USA

\* Corresponding author: raymer@uoregon.edu

June 9, 2021

## Abstract:

Two-photon absorption (TPA) and other nonlinear interactions of molecules with time-frequency-entangled photon pairs (EPP) has been predicted to display a variety of fascinating effects. Therefore, their potential use in practical quantum-enhanced molecular spectroscopy requires close examination. This paper presents in tutorial style a detailed theoretical study of one- and two-photon absorption by molecules, focusing on how to treat the quantum nature of light. We review some basic quantum optics theory, then we review the density-matrix (Liouville) derivation of molecular optical response, emphasizing how to incorporate quantum states of light into the treatment. For illustration we treat in detail the TPA of photon pairs created by spontaneous parametric down conversion, with an emphasis on how quantum light TPA differs from that with classical light. In particular, we treat the question of how much enhancement of the TPA rate can be achieved using entangled states. The paper includes review of known theoretical methods and results, as well as some extensions, especially the comparison of TPA processes that occur via far-off-resonant intermediate states only and those that involve off-resonant intermediate state by virtue of dephasing processes. A brief discussion of the main challenges facing experimental studies of ETPA is also given.

## TABLE OF CONTENTS

## 1. Introduction

In the past decade many theoretical studies have proposed that 'quantum advantages' in spectroscopy can be achieved by the use of time-frequency-entangled photon pairs (EPP). For review see [1,2,3,4] These include, for example, proposals for virtual-state spectroscopy [5,6], Raman spectroscopy [7], and multi-dimensional optical spectroscopy [8,9]. Two-photon absorption (TPA) with entangled light can provide a testbed for many of these ideas, as indicated by a flurry of recent papers addressing the experimental feasibility of using entangled two-photon absorption (ETPA) to increase the measured signal in spectroscopic or imaging applications. This paper presents not a complete review, but a tutorial treatment of the underlying quantum optics needed for understanding how quantum light interacts with molecules, especially one- and two-photon absorption. We review some basic quantum optics theory, then we review the density-matrix (Liouville) derivation of molecular optical response, emphasizing how to incorporate quantum states of light into the treatment. The treatment focuses on photon pairs at very low flux where different pairs do not overlap, and addresses in detail the question of the degree of enhancement of TPA by quantum correlations of photon number and in spectral properties. The paper includes review of known theoretical methods and results, as well as some extensions, especially the comparison of TPA processes that occur via far-off-resonant intermediate states only and those that involve off-resonant intermediate state by virtue of dephasing processes. A brief discussion of the main challenges facing experimental studies of ETPA is also given.

We consider the three well-known 'quantum pathways' in the fourth-order perturbation theory of the density matrix, and clarify to what extent the TPA probability can be modelled by the conventional second-order perturbation theory using state amplitudes. The treatment contains results not previously published, to our knowledge, regarding the roles of the coherent excitation of multiple nonresonant intermediate states. These pathways are known as 'step-wise' pathways or 'non-rephasing' and 'rephasing' pathways, whereas the conventional second-order perturbation theory includes only the direct 'two-quantum' or 'double quantum coherence' pathway that proceeds through a set of nonresonant virtual states. The two-quantum pathway is known to be enhanced by frequency anticorrelations of the exciting photons. A major question that we address is whether the step-wise pathways are similarly enhanced. We find that they are not.

The basis of the current thinking on entangled two-photon absorption goes back to the seminal theoretical papers by Klyshko [10], Gea-Banacloche [11] and Javanainen and Gould. [12] These initial studies showed that in the regime of 'isolated' EPP—defined as the regime of extremely low flux where not more than two photons on average impinge on the molecule within the field's coherence time and not more than two photons impinge on the molecule within its electronic dephasing time — the TPA probability for exciting a molecule scales linearly with the photon flux because the photons arrive in pairs. Moreover, for EPP created by spontaneous parametric down conversion (SPDC) pumped by a narrow-band laser, an increased bandwidth of the EPP field does not decrease the absorption probability. This effect occurs because the frequencies of the two EPP photons are not random, but rather anticorrelated so that they sum to the fixed frequency of the pump laser. We refer to this effect as enhancement due to spectral correlation. The first of these predictions was verified in an experiment using narrow-band EPP to excite a two-photon transition in a vapor of atomic cesium. [13] The second prediction found experimental support in atomic systems for a slightly different scenario: TPA in the high flux (squeezing) regime where the linear scaling of the TPA rate is lost, but the enhancement by frequency anticorrelation is retained. [14] Subsequent studies using molecular solutions reported many orders-of-magnitude enhancement of TPA using EPP [15, 16], but those studies have been called into question by recent experimental studies. [17, 18, 19]

Another issue is whether or not a *single-photon* state excites a molecule any differently than does a weak coherent state. Given that a one-photon transition of a single molecule can absorb only one photon at a time, does the molecule 'know' what state of light that photon is provided by? The short answer is 'no.' But when more than one molecule is present, even in the absence of direct Coulomb coupling between molecules, quantum entanglement between molecules can occur as a result of their being coupled to the same optical field.

We first introduce the overall problem of one- and two-photon absorption of quantum light. We then review the formalism of the quantized electromagnetic field. We then find and apply perturbative solutions to the density matrix equations of motion to derive excitation probabilities for one-photon and two-photon absorption, emphasizing differences between classical (coherent-state) excitation and quantum (isolated EPP) excitation. We review and extend a recent derivation predicting the amount of quantum enhancement of TPA that can be achieved using isolated EPP compared to the same flux of classical light, from which it is concluded that observation of two-photon excitation of molecules by isolated EPP is challenging in practice. [20, 21, 22]

Dispersion of the EPP field by passage through typical linear-optical elements is known to decrease the efficacy of TPA by EPP. Our formalism allows incorporating rigorously the effects of dispersion in TPA, leading to a simple formula quantifying the expected decrease of the EPP-driven TPA probability.

Several theoretical studies with goals similar to ours have appeared previously but considered only a subset of the issues we treat. Dayan emphasized cases with no resonant intermediate states, which is also our focus, and also treated the case of overlapping EPP at high flux. [23] Schlawin and Buchleitner emphasized cases in which resonant intermediate states play a strong role in TPA. [24]

## 2. Theory of one- and two-photon absorption—General formalism

The theory of interaction of molecules with quantized light, including one- and two-photon absorption, has been treated numerous times. The traditional approach, first developed by Maria Göppert-Mayer, uses perturbation theory for state amplitudes and posits a density of molecular or photonic states to arrive at the Fermi Golden Rule for the *conventional cross section* for TPA. [25, 26, 27] Accessible textbook treatments are given in the quantum optics text by Loudon [28] and in the nonlinear optics text by Boyd. [29] When dealing with short light pulses or light having time-frequency correlations (entanglement), a more detailed treatment is needed, and several such treatments have appeared, a few being [1, 11, 12, 20, 21]

**Figures 1** and **2** define the molecular states of interest and the perturbative sequences for the molecular density-matrix elements leading to excitation by one- or two-photon transitions. Molecular density-matrix elements are given by the total density operator $\hat{\rho}$ as:

$$\rho_{ij} = Tr\left(\hat{\rho}|j\rangle\langle i|\right), \tag{1}$$

where $|j\rangle$ is a molecular energy eigenstate ('ket') and $\langle j|$ is its corresponding 'bra'. The arrows in the figures correspond to single interactions with the field. Note that a single interaction creates a 'coherence' between two states, while a second interaction is needed to create a population. To clarify: If there is no dipole dephasing present, a 'population' $\rho_{ee}$ is simply the square of a 'coherence' $\rho_{eg}$, the squaring process corresponds to a second interaction as in **Figs. 2e** and **f**. We will return to these diagrams repeatedly.

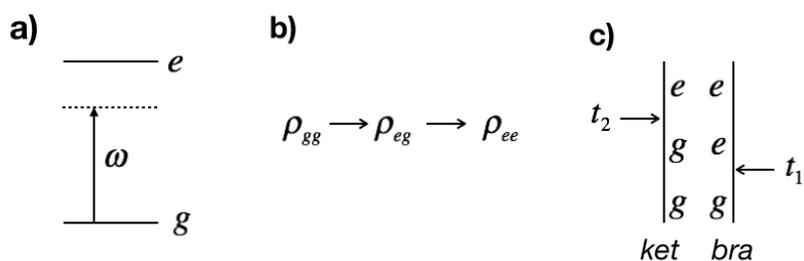

**Fig. 1** a) A molecule (or atom, etc.) with ground and excited states labeled $g$ and $e$, with energies $E_i = \hbar\omega_i$. b) The perturbative sequence in which molecular density-matrix elements $\rho_{ij}$ are excited. c) Double-sided Feynman diagram representing the sequence of field interactions needed to create a 'coherence' (one interaction) or a 'population' (two interactions).

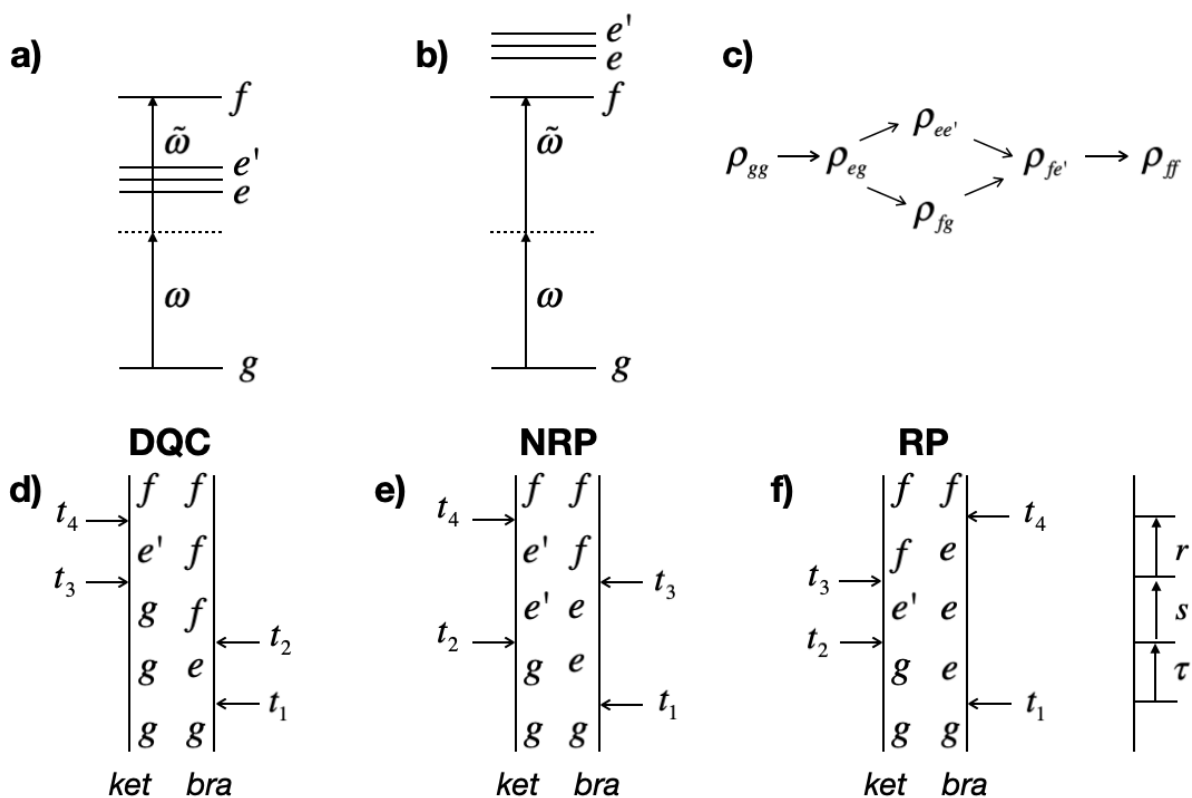

**Fig. 2** a) A molecule with ground, intermediate, and final states labeled $g$, $e$ or $e'$, and $f$, with energies $E_i = \hbar\omega_i$. b) A case where the intermediate states are higher in energy than the final state. c) The perturbative sequence in which molecular density-matrix elements are excited. d-f) Double-sided Feynman diagram representing the three possible pathways. d) *Double quantum coherence* (DQC) pathway, e) *Non-rephasing* (NRP) pathway, and f) *Rephasing* (RP) pathway. The NRP and RP pathways include intermediate-state coherences (*e,e'*) and/or populations (*e,e*), while DQC does not involve the (*e,e'*) or (*e,e*) density-matrix elements. The difference-time variables $\tau$, *s* and *r* are indicated to the far right.

We follow, as far as possible, the standard literature in ultrafast spectroscopy, where it is common to use a molecular density-matrix formalism to model absorption or nonlinear-optical response of multilevel systems. [1, 2, 3, 4, [30]] In a recent paper [20] we showed how to obtain the correct results for the DQC pathway of **Fig. 2d** using an alternative formalism that is common in the quantum optics literature – the 'operator optical Bloch equations'. [[31]] These equations respect the quantum nature of the field via commutation relations involving raising and lowering operators for the field and the molecular states. However, for treating the NRP and RP pathways of **Fig. 2e** and **f**, the operator optical Bloch equations become cumbersome, and the density matrix method appears to be more straightforward.

Consider a single molecule interacting with a quantum light field. The molecular energy eigenstates satisfy the eigen-equation, $\hat{H}_M |i\rangle = \hbar\omega_i |i\rangle$. The density matrix equations track the time evolution of molecular-state density-matrix elements $\rho_{ji}$, which equate to populations (that is probabilities) for $i = j$ and 'coherences' for $i \neq j$. In the Schroedinger Picture, the density matrix equation of motion for the combined molecule-field system (without damping or dephasing interactions) is $\partial_t \hat{\rho} = (1/i\hbar)[\hat{H}, \hat{\rho}]$, where the Hamiltonian is $\hat{H} = \hat{H}_0 - \vec{d} \cdot \Sigma_\sigma \mathbf{e}^\sigma \hat{E}^\sigma$, and the unperturbed Hamiltonian is $\hat{H}_0 = \hat{H}_M + \hat{H}_F$, with $\hat{H}_F$ being the energy of the field; $\vec{d}$ is the electric-dipole-vector operator, and $\mathbf{e}^\sigma \hat{E}^\sigma$ is the field operator with $\mathbf{e}^\sigma$ being its linear polarization vector, where the sum over $\sigma$ takes on two values ($\sigma = 1, 2$).

The electric field operator, at the location of the molecule, is written as $\hat{E}^\sigma(t) = \hat{E}^{(+)}(t) + \hat{E}^{(-)}(t)$, where $\hat{E}^{(-)}(t)$ is a wide-band (all frequencies) photon creation operator and $\hat{E}^{(+)}(t)$ is a wide-band annihilation operator. In the semiclassical approximation one assumes the field, with state $\hat{\rho}_F$, is not quantum correlated with the molecular response, described by the state $\hat{\rho}_M$, so the combined state is the tensor product $\hat{\rho}(t) = \hat{\rho}_M(t) \otimes \hat{\rho}_F(t)$. Then the mean-field approximation corresponds to the replacement $Tr(\hat{\rho} |j\rangle\langle i| \hat{E}^\sigma) \rightarrow Tr(\hat{\rho}_M |j\rangle\langle i|) Tr(\hat{\rho}_F \hat{E}^\sigma) = \rho_{ij} \langle \hat{E}^\sigma \rangle$, which recovers the standard molecular density matrix treatment (standard optical Bloch equations) with the field operator replaced by the mean field. This approximation precludes treatment of quantum states of light, which are characterized by the correlation functions $\langle \hat{E}(t_1) \hat{E}(t_2) \hat{E}(t_3) ... \rangle$. Therefore, in order to allow for the possibility that the field is quantum correlated with the molecular response, we continue to treat the field as an operator.

It is easiest to express the solutions for the density operator in the Interaction Picture (indicated by $I$ subscripts), in which observables are expressed as $\hat{O}_I(t) = \hat{U}_0^{\dagger}(t,t_0)\hat{O}\hat{U}_0(t,t_0)$, where $\hat{U}_0(t,t_0) = \exp\left(-i\hat{H}_0(t-t_0)\right)$, with $t_0$ being an arbitrary time at which the Interaction Picture and Schroedinger Picture are equivalent. We set $t_0 = 0$ for convenience. The density operator in the new picture is:

$$\hat{\rho}_I(t) = \hat{U}_0^{\dagger}(t,t_0)\hat{\rho}(t)\hat{U}_0(t,t_0) \ , \tag{2}$$

and system operators in this picture are:

$$\hat{O}_I(t) = \hat{U}_0^{\dagger}\hat{O}\hat{U}_0 \ . \tag{3}$$

where we suppressed $(t,t_0)$. The expectation value of an observable is expressed as (using $\hat{U}_0\hat{U}_0^{\dagger} = \hat{1}$):

$$
\begin{aligned}
\left\langle \hat{O} \right\rangle &= Tr\left(\hat{\rho}(t)\hat{O}\right) \\
&= Tr\left(\hat{U}_0\hat{U}_0^{\dagger}\hat{\rho}(t)\hat{U}_0\hat{U}_0^{\dagger}\hat{O}\right) \\
&= Tr\left(\hat{U}_0^{\dagger}\hat{\rho}(t)\hat{U}_0\hat{U}_0^{\dagger}\hat{O}\hat{U}_0\right) \\
&= Tr\left(\hat{\rho}_I(t)\hat{O}_I(t)\right),
\end{aligned}
\tag{4}
$$

where we used cyclic permutation under the trace and suppressed the time arguments of $\hat{U}_0(t,t_0)$ for clarity.

The Interaction-Picture density operator satisfies:

$$\partial_t\hat{\rho}_I(t) = (1/i\hbar)[\hat{H}_I(t),\hat{\rho}_I(t)] \ , \tag{5}$$

in which the interaction Hamiltonian is:

$$\hat{H}_I(t) = -\hat{d}_I(t)\hat{E}_I(t) \ . \tag{6}$$

The perturbative solution for the density operator is given by the series: [30, [32]]

$$\hat{\rho}_I(t) = \hat{\rho}_0 + \sum_{n=1}^{\infty}\hat{\rho}_I^{(n)}(t) \ , \tag{7}$$

where $\hat{\rho}_0 = \hat{\rho}_M(0)\otimes\hat{\rho}_F(0)$ is the initial state of the system, assumed to be uncorrelated long before the interaction begins ($t = -\infty$), and the $n$th iterate is:

$$\hat{\rho}_I^{(n)}(t) = \left(\frac{1}{i\hbar}\right)^n \int_{-\infty}^{t} dt_n \int_{-\infty}^{t_n} dt_{n-1} ... \int_{-\infty}^{t_2} dt_1 \left[\hat{H}_I(t_n), \left[\hat{H}_I(t_{n-1}), ... \left[\hat{H}_I(t_1), \hat{\rho}_0\right]\right]\right]. \tag{8}$$

It is worth noting that we have not set $t_0$ to be the same as the time at which the initial state is specified, thereby letting us simplify the equations.

The ket-bra operators $|i\rangle\langle j|$ (like Pauli matrices) can be used to represent any molecular operator, such as the electric dipole operator, $\hat{d}_I(t) = \hat{d}_I^{(-)}(t) + \hat{d}_I^{(+)}(t)$, where (setting $t_0 = 0$):

$$\begin{aligned} \text{raising:} \quad & \hat{d}_I^{(-)}(t) = \sum_{j, i>j} d_{ij} |i\rangle\langle j| e^{i(\omega_i - \omega_j)t} \\ \text{lowering:} \quad & \hat{d}_I^{(+)}(t) = \sum_{j, i<j} d_{ij} |i\rangle\langle j| e^{i(\omega_i - \omega_j)t} \end{aligned}, \tag{9}$$

and the electric-dipole matrix elements are $d_{ij} = \langle i|\vec{d}\cdot\mathbf{e}^\sigma|j\rangle$. These operators act as molecular raising and lowering operators. Note that Interaction-Picture operators referring to different degrees of freedom (such as molecule and field) commute, even at different times, because they equal Schroedinger-Picture operators multiplied by complex-valued (non-operator) functions of time.

In order to identify clearly the various quantum pathways leading to the molecular populations and coherences of interest, we will analyze the perturbative solutions of the equations of motion. Before we do that, we review the quantum theory of light.

## 3. Quantization of traveling-wave optical fields

How does light travel through vacuum? At every point, even in vacuum, there exist an infinite number of electromagnetic harmonic oscillators, one for each of a continuum of frequencies, which are coupled via the Maxwell equations to their same-frequency neighbors, creating propagation. The optical 'modes' of the system are the collective degrees of freedom of all the oscillators that form natural solutions of the Maxwell Equations.

The electric field $\mathbf{E}(\mathbf{r})$ represents the summed amplitude of all the oscillators at point $\mathbf{r}$. In this view, a 'photon' is not a tangible object. The word photon is used merely to name various states of the field: a one-photon state of the field, a two-photon state of the field, etc. This viewpoint, developed by Dirac, is the most common one used in quantum optics, although you will often hear the word photon bandied about in not-so-cautious ways.

In the quantum theory of a harmonic oscillator (e.g., a mass on a spring), the raising operator $\hat{a}^\dagger$ increases the energy by one unit, given by $\hbar\omega_0$, where $\omega_0$ is the angular frequency of the oscillator. In the quantum theory of light, a given raising operator $\hat{a}_j^\dagger$ increases the excitation number in the corresponding mode by one: $n \to n+1$. If a raising operator acts on the lowest, ground state of the system (the 'vacuum'), a one-photon state of the field is created. A raising operator $\hat{a}_j^\dagger$ is also called a 'creation operator,' not because it creates an object called a photon but because when acting on the vacuum it creates an excitation of the field, described by a state given the name 'photon.' That is, a photon is best thought of as a state of the field, not as a particle. [33,34]

Many molecular spectroscopic studies are carried out in a liquid or solid host medium. Strictly speaking, excitations of the electromagnetic (EM) field inside a medium should be called polaritons, not photons, because they are collective excitations of the field and electronic states of the medium together. [35,36] For ease of language we will continue to use the word photon.

Notice that a one-photon state of the field need not be monochromatic because it may be spread over many spectral modes in a coherent quantum superposition. [37] For example, if the state is generated by spontaneous emission by an excited-state atom, the emitted light is a wave packet and its spectrum is that of the natural linewidth associated with the lifetime of the excited state. Thus, we will discuss in this tutorial how to quantize the EM field in terms of nonmonochromatic modes, also called temporal modes or pulse modes. [38, 39]

In quantum electromagnetism, the infinite set of raising and lowering operators together are used to construct a global operator $\hat{\mathbf{E}}(\mathbf{r})$. In the Heisenberg picture, where operators evolve in time and states do not, this operator can be expressed in terms of monochromatic modes $\tilde{\mathbf{u}}_j(\mathbf{r})$, each having a particular frequency $\omega_j$. Here we focus on cases where propagation occurs approximately in one dimension only ($z$), such as light in a well-collimated beam or in an optical wave guide such as a fiber. We can choose a set of monochromatic beam-like mode functions as $\mathbf{e}_j \mathrm{w}_j(\mathbf{x},z,\beta_j)\exp(i\beta_j z)$, where $\mathbf{e}_j$ is one of two polarization unit vectors perpendicular to the direction of propagation. The $\mathrm{w}_j(\mathbf{x},z,\beta_j)$ are transverse modes that depend strongly on $\mathbf{x}=(x,y)$ and weakly or not at all on $z$. They are orthogonal and square normalized in the transverse variable $\mathbf{x}$. Here $\beta_j$ is the longitudinal propagation constant (wave number); periodic boundary conditions require it to have equally spaced values $\beta_j = j2\pi/L$, where $j$ is any integer, typically of order $10^6$ for optical fields. For each value of the longitudinal propagation constant $\beta_j$ there exists a discrete set of transverse modes, which generally have different frequencies depending on their transverse spatial shape. If the medium is only weakly dispersive, the transverse modes with equal $\beta$ are orthonormal in two-dimensions (for any $z$ value) according to:

$$\int d^2x \, \mathrm{w}^*_i(\mathbf{x},\beta)\mathrm{w}_j(\mathbf{x},\beta) = \delta_{ij} \; . \tag{10}$$

When propagation is quasi-one-dimensional, it is useful to adopt angular frequency $\omega$ as the continuous variable used to label modes, since we typically measure frequency. Note that for each transverse mode $j$ there is a dispersion relation $\omega_j(\beta)$, determined by the solutions of the Maxwell Equations (for example, the Hermite-Gaussian modes propagation in a uniform medium). Considering the optical spectrum in free space as a continuum, the field operator describing a not-too-broad spectral band of light is $\hat{\mathbf{E}} = \hat{\mathbf{E}}^{(+)} + \hat{\mathbf{E}}^{(-)}$, with: [40, 36]

$$\hat{\mathbf{E}}^{(+)}(\mathbf{r},t) = i \sum_j \int_{-\infty}^{\infty} \frac{d\omega}{2\pi} \sqrt{\frac{\hbar\omega}{2\varepsilon_0 c\, n}} \hat{a}_j(\omega) e^{-i\omega t} \mathbf{e}_j \, \mathrm{w}_j(\mathbf{x},z,\omega) \exp[i\beta_j(\omega)z] \ . \tag{11}$$

The frequency $\omega$ has units rad/s, $n$ is the medium's refractive index at the central frequency of the spectral band of the field, $\varepsilon_0$ is the vacuum permittivity, and $c$ is the vacuum speed of light. The dispersion relation is now written $\beta_j(\omega)$. The continuum operators satisfy the commutator:

$$\left[ \hat{a}_j(\omega), \hat{a}_j^\dagger(\omega') \right] = 2\pi \delta(\omega - \omega') \delta_{ij} \ . \tag{12}$$

The so-called negative-frequency part $\hat{\mathbf{E}}^{(-)}$ is related to the positive-frequency part $\hat{\mathbf{E}}^{(+)}$ by an operator conjugate: $\hat{\mathbf{E}}^{(-)} = \hat{\mathbf{E}}^{(+)\dagger}$. If fields of very different central frequencies are considered, then Eq.(11) needs to be written for each spectral band separately.

Now consider the operator for the field of a single spatial-polarization mode at the location $(\mathbf{x}_M, z_M)$ of a molecule. Dropping the mode index $j$, and replacing the frequency by its the central (mean) value $\bar{\omega}$, Eq.(11) becomes (we suppress the $I$ subscript indicating the Interaction Picture):

$$\hat{\mathbf{E}}^{(+)}(\mathbf{r}_M, t) = i e^{i\theta} \mathbf{e}\, \hat{E}^{(+)}(t) \ , \tag{13}$$

where the scalar part of the electric field operator is:

$$\hat{E}^{(+)}(t) = \int \mathrm{d}\omega\, L(\omega) \hat{a}(\omega) e^{-i\omega t} \ , \tag{14}$$

and $\hat{E}(t) = \hat{E}^{(+)}(t) + \hat{E}^{(-)}(t)$, with $\hat{E}^{(-)} = \hat{E}^{(+)\dagger}$ and $L(\omega) = (\hbar\omega / 2\varepsilon_0 nc)^{1/2}\, \mathrm{w}(\mathbf{x}_M, z_M, \omega)$, where $\mathrm{w}(\mathbf{x}_M, z_M, \omega)$ is the mode amplitude at the molecule's location. The phase factor is given by $\theta = \beta(\bar{\omega})z_M$ and for simplicity we can set this to an arbitrary value if only one molecule is being considered. Here and in the following we use the shorthand notation $\mathrm{d}\omega = d\omega / 2\pi$.

We typically consider a state of the field that has a bandwidth much narrower than its central frequency $\omega_0$. Then we approximate the field operator as:

$$\hat{E}^{(+)}(t) = L_0 \int \mathrm{d}\omega\, \hat{a}(\omega) e^{-i\omega t}, \tag{15}$$

where $L_0 = (\hbar\omega_0 / 2\varepsilon_0 ncA_0)^{1/2}$, and $A_0$ is the *effective* beam area at the molecule's location given by $A_0^{-1/2} \equiv \mathrm{w}(\mathbf{x}_M, z_M, \bar{\omega})$. In this approximation, the adjoint operator $\hat{E}^{(-)}(t)$ creates a photon at time $t$ at exactly the location $(\mathbf{x}_M, z_M)$.

Strictly speaking, the formalism just given is valid only if the medium is nearly transparent in the spectral range of interest, so only dispersion affects the light. [35, 36] In the present case, we consider only one molecule as an absorber and consider that the host medium is transparent to the light, so the above formalism holds.

## 4. Classical and quantum states of light

Here we summarize the basic properties of coherent states, single-photon states, and two-photon states. For convenience, we consider the incident light to be in pulses of finite duration. In the case of continuous-wave (CW) excitation, we imagine the field to be made of a series of rectangular pulses with constant mean power and duration $T$, as in **Fig. 3(a)**. When comparing to TPA with short pulses of SPDC or coherent-state light, it suffices to consider only a single pulse occupying the same interaction time window $T$, as in **Figs. 3(b) and (c)**.

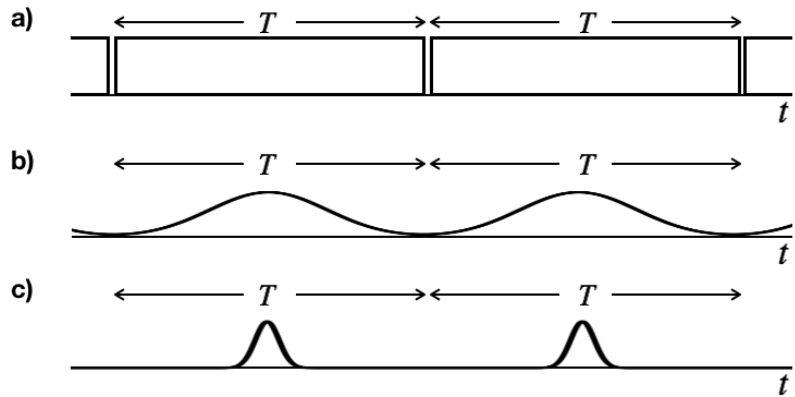

Fig. 3 Rectangular (a) and Gaussian pulse trains (b and c) with period $T$.

### 4.1. Coherent state

A pulsed coherent state $|\alpha\rangle$ with spectral amplitude $\alpha(\omega)$ satisfies $\hat{a}(\omega)|\alpha\rangle = \alpha(\omega)|\alpha\rangle$ at each frequency, [40] or equivalently in the time domain:

$$\hat{E}^{(+)}(t)|\alpha\rangle = L_0 A(t)e^{-i\omega_0 t}|\alpha\rangle, \tag{16}$$

where the field amplitude is:

$$A(t) = \int d\omega\, \alpha(\omega)e^{-i(\omega-\omega_0)t}\ . \tag{17}$$

$A(t)$ is a slowly varying envelope and $\omega_0$ is the central (carrier) frequency. Because the coherent state is the quantum state that most resembles a field in classical EM theory, we often call the coherent state a 'classical' state.

The mean number of photons in the pulse is the time-integral of the flux $|A(t)|^2$:

$$N = \int |\alpha(\omega)|^2 d\omega = \int |A(t)|^2\, dt\ . \tag{18}$$

It is useful to define a unity-normalized spectral amplitude $\phi(\omega)$ by $\alpha(\omega) = \alpha_0 \phi(\omega)$, where $\int |\phi(\omega)|^2 \, d\omega = 1$. Then $N = |\alpha_0|^2$.

In the case of a coherent state in a constant-amplitude ('rectangular') pulse with duration $T$, the needed Fourier-transform pair is:

$$A(t) = \{\alpha_0 T^{-1/2}, |t| < T/2 \; ; \; 0 \; else\}$$

$$\phi(\omega) = \frac{\sin[(\omega - \omega_0)T/2]}{\sqrt{T}(\omega - \omega_0)/2} \qquad , \qquad (19)$$

as illustrated in **Fig. 4**.

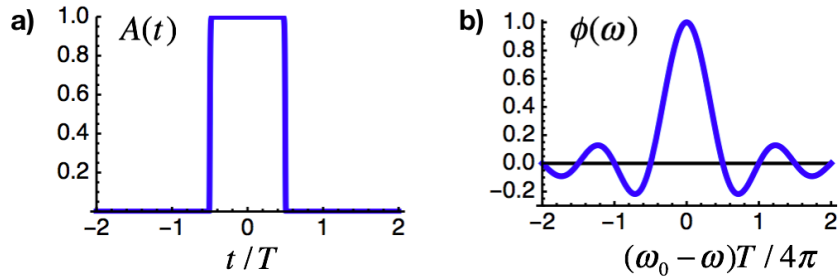

Fig.4 a) Rectangular coherent-state amplitude vs. time, b) corresponding spectral amplitude, both for $T = 1$ and $\alpha_0 = 1$.

### 4.2. Single-photon state

A single-photon state of a particular temporal mode (coherent wave packet) $\varphi(\omega)$ is described by a superposition of monochromatic one-photon states:

$$|\varphi\rangle = \int d\omega \, \varphi^*(\omega) \hat{a}^\dagger(\omega) |vac\rangle , \qquad (20)$$

where the spectral density (probability) is normalized as $\int d\omega |\varphi(\omega)|^2 = 1$. This state corresponds to light arriving at the molecule in the form of a time-domain wave packet:

$$\tilde{\varphi}(t) \equiv \int d\omega \, \varphi(\omega) e^{-i\omega t} . \qquad (21)$$

Such a state can be created by, for example, heralded SPDC in which one of the photons is detected, announcing or 'heralding' the presence of the other. [41, 42].

It is sometimes useful to quantize the field in terms of wave packets rather than monochromatic waves. [37, 38, 39, 40] This can be done by choosing any complete orthonormal set of spectral amplitude functions $\{\varphi_j(\omega)\}$, with $\sum_j \varphi_j^*(\omega')\varphi_j(\omega) = 2\pi\delta(\omega'-\omega)$, and using them to define a set of 'wave packet operators':

$$\hat{a}_j = \int \!\!d\!\!\!\!/\omega\, \varphi_j^*(\omega)\hat{a}(\omega). \tag{22}$$

These operators satisfy $[\hat{a}_i, \hat{a}_j^\dagger] = \delta_{ij}$. The continuous set of operators has been converted into a discrete set. One can create a single-photon state of the field by acting such an operator on the vacuum state: $\hat{a}_j^\dagger |vac\rangle$. This action corresponds to creating a photon in a particular 'temporal mode,' as can be seen by inverting Eq.(22):

$$\hat{a}(\omega) = \sum_j \varphi_j(\omega)\hat{a}_j \ . \tag{23}$$

So, the electric field operator Eq.(15) can be expressed as:

$$\hat{E}^{(+)}(t) = L_0 \sum_j \hat{a}_j \mathrm{v}_j(t), \tag{24}$$

where $\mathrm{v}_j(t)$ is called a temporal mode:

$$\mathrm{v}_j(t) = \int \!\!d\!\!\!\!/\omega\, \varphi_j(\omega)e^{-i\omega t} \ . \tag{25}$$

A useful example of a square-normalized temporal mode spectrum is given by:

$$\varphi(\omega) = \frac{\sqrt{2\Gamma}}{\Gamma - i(\omega - \omega_0)} \ , \tag{26}$$

where $\Gamma$ is the spectral half-width. Then the temporal mode is a single-sided exponential:

$$\mathrm{v}(t) = \begin{cases} 0, \ t < 0 \\ \sqrt{2\Gamma}\exp[-(i\omega_0 + \Gamma)t] \, , t > 0 \end{cases} \ , \tag{27}$$

as illustrated in **Fig. 5**.

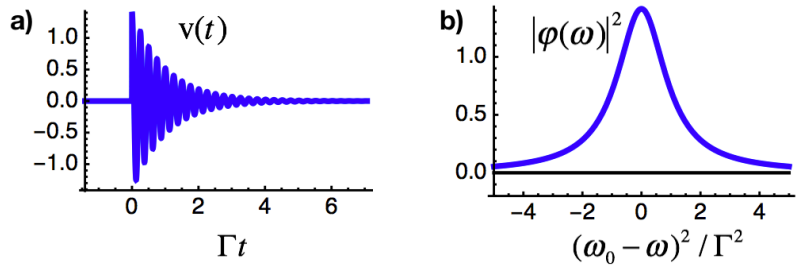

Fig. 5 Example of a temporal mode, a) in time, and b) in frequency, for $\Gamma = 1$.

### 4.3. Entangled photon pair state

Collinear Type-0 or Type-I spontaneous parametric down conversion is characterized by the two generated photons having the same polarization. When pumped by a laser pulse of finite duration the SPDC can be designed to occur into a single spatial-and-polarization mode [43]; then the state is described by:

$$|\Psi\rangle = \sqrt{1-\varepsilon^2}\,|vac\rangle + \varepsilon \int d\omega \int d\tilde{\omega}\,\psi(\omega,\tilde{\omega})\hat{a}^\dagger(\omega)\hat{a}^\dagger(\tilde{\omega})|vac\rangle + ... \,, \quad (28)$$

where $\varepsilon^2 << 1$ is the probability that a given pulse contains a photon pair. Consistent with the labeling of pulses shown in **Fig. 3**, the mean photon flux (twice the pairs flux), time-averaged over the interval $T$, equal $F_{EPP} = 2\varepsilon^2/T$, for either CW or pulsed cases. We neglect higher-order terms representing generation of multiple pairs in order to satisfy our assumption of an isolated EPP interacting with the molecule. The joint-spectral amplitude (JSA) $\psi(\omega,\tilde{\omega})$ is determined by energy and momentum conservation, given the spectrum of the pumping laser pulse and the phase-matching properties of the nonlinear crystal used as second-order nonlinear medium. [44] It is normalized as $\int d\omega \int d\tilde{\omega} \,|\psi(\omega,\tilde{\omega})|^2 = 1$. An important property of the JSA when treating a single mode of the field, for which there are no distinguishing labels on the photon creation operators other than frequency, is the required symmetry $\psi(\omega,\tilde{\omega}) = \psi(\tilde{\omega},\omega)$. This fact is seen easily by swapping the integration variables in Eq.(28). Such a symmetry is not upheld when treating Type-II SPDC, which takes place into two or more distinct modes.

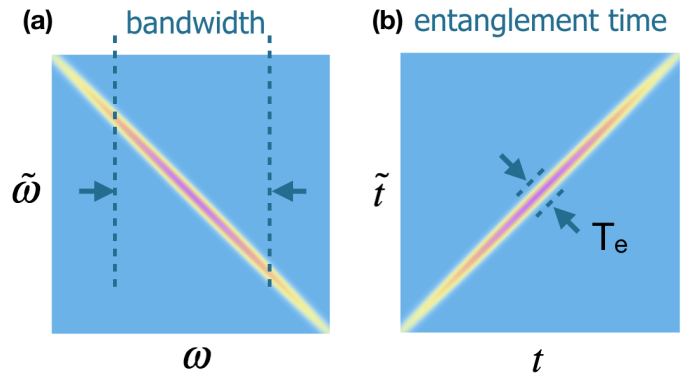

Fig. 6 Example of the joint-spectral amplitude (JSA) of a two-photon state, (a) in the frequency domain $\psi(\omega,\tilde{\omega})$, and (b) the same state in the time domain $\tilde{\psi}(t,\tilde{t})$.

For a long narrowband pump pulse with central frequency $\omega_P$, the JSA is largest typically along he antidiagonal, that is $\omega + \tilde{\omega} = \omega_P$, where the frequencies are anti-correlated, as illustrated in **Fig. 6a**. The two-photon wave packet in the time domain is given by the double Fourier transform:

$$\tilde{\psi}(t,\tilde{t}) = \int d\tilde{\omega} \int d\omega\, e^{-i(\tilde{\omega}-\omega_0)\tilde{t}}\,e^{-i(\omega-\omega_0)t}\,\psi(\omega,\tilde{\omega}) \,. \quad (29)$$

The photon arrival times are positively correlated to within the so-called *entanglement time* $T_e$, as shown in **Fig. 6b**. The entanglement time is given roughly by the inverse of the bandwidth.

Because Eq.(28) represents the state of a single spatial mode, it does not describe transverse spatial entanglement (correlation). The photons, if detected, are assumed to be distributed across the beam independently, as is the case if their generation takes place in a single-mode wave guide or the beam is spatially filtered by passing through a single-mode optical fiber. Therefore, in this case with EPP generated by SPDC, the 'entanglement area' is the transverse area of the EPP beam at the molecule's location, unlike in treatments where transverse spatial correlations are considered. [5] For a wide-area beam, such correlations can localize photon pairs to transverse areas much smaller than the overall beam area but cannot localize pairs to an area much smaller than the diffraction-limited focus of a well-designed optical system.

## 5. Induced dipole, one-photon absorption, and dephasing

### 5.1. First-order solution

Here we give the details of the lowest-order solution of the density-matrix and its application to the creation of optical coherence and one-photon absorption by a single-photon wave packet. For simplicity, again we assume the field has energy in only one polarization state and drop the $\sigma$ index.

The first-order solution for the density matrix describes the coherent linear response of a molecule:

$$\hat{\rho}_I^{(1)}(t) = \frac{1}{i\hbar} \int_{-\infty}^{t} dt_1 \left[ \hat{H}_I(t_1), \hat{\rho}_0 \right].$$ (30)

A useful step in formulating the solutions is to transform to a difference-time variable, $\tau = t - t_1$. Then the solution is:

$$\hat{\rho}_I^{(1)}(t) = \frac{1}{i\hbar} \int_0^\infty d\tau \left[ \hat{H}_I(t-\tau), \hat{\rho}_0 \right]$$

$$= \frac{1}{i\hbar} \int_0^\infty d\tau \hat{H}_I(t-\tau)\hat{\rho}_0 \; + \; hc \qquad , \qquad (31)$$

$$= \frac{-1}{i\hbar} \int_0^\infty d\tau \, \hat{d}_I(t-\tau)\hat{E}_I(t-\tau)\hat{\rho}_0 \; + \; hc$$

where *hc* stands for Hermetian conjugate, consistent with the requirement that the density operator be Hermetian.

The macroscopic electric-dipole polarization density is equal to the number density of molecules times the expectation of the dipole operator, given by $\left\langle \hat{a}_I(t) \right\rangle = \left\langle \hat{a}_I^{(-)}(t) \right\rangle + cc$, where, using $\hat{\rho}_0 = \hat{\rho}_M \hat{\rho}_F$ and permutation inside the trace:

$$
\begin{aligned}
\left\langle \hat{a}_I^{(-)}(t) \right\rangle &= Tr\left( \hat{\rho}_I(t) \hat{a}_I^{(-)}(t) \right) \\
&= Tr\left( \hat{a}_I^{(-)}(t) \frac{-1}{i\hbar} \int_0^\infty d\tau \left[ \hat{a}_I(t-\tau)\hat{E}_I(t-\tau), \hat{\rho}_M \hat{\rho}_F \right] \right) \\
&= \frac{-1}{i\hbar} \int_0^\infty d\tau \left\langle \hat{a}_I^{(-)}(t)\hat{a}_I(t-\tau) \right\rangle \left\langle \hat{E}_I(t-\tau) \right\rangle + \frac{1}{i\hbar} \int_0^\infty d\tau \left\langle \hat{a}_I(t-\tau)\hat{a}_I^{(-)}(t) \right\rangle \left\langle \hat{E}_I(t-\tau) \right\rangle \ ,
\end{aligned}
\tag{32}
$$

where:

$$
\begin{aligned}
\left\langle \hat{a}_I^{(-)}(t)\hat{a}_I(t-\tau) \right\rangle &= Tr_M\left( \hat{\rho}_M \hat{a}_I^{(-)}(t)\hat{a}_I(t-\tau) \right) \\
\left\langle \hat{a}_I(t-\tau)\hat{a}_I^{(-)}(t) \right\rangle &= Tr_M\left( \hat{\rho}_M \hat{a}_I(t-\tau)\hat{a}_I^{(-)}(t) \right) \\
\left\langle \hat{E}_I(t-\tau) \right\rangle &= Tr_F\left( \hat{\rho}_F \hat{E}_I(t-\tau) \right) \ .
\end{aligned}
\tag{33}
$$

Throughout this tutorial we assume the molecule starts in its ground state, so $\hat{\rho}_M = \left| g \right\rangle \left\langle g \right|$. Then, as a consequence of $\left\langle g \right| \hat{a}_I^{(-)}(t) = 0$:

$$
\begin{aligned}
\left\langle \hat{a}_I^{(-)}(t)\hat{a}_I(t-\tau) \right\rangle &= \sum_j \left\langle j \right| \left( \left| g \right\rangle \left\langle g \right| \hat{a}_I^{(-)}(t)\hat{a}_I(t-\tau) \right) \left| j \right\rangle \\
&= 0 \ ,
\end{aligned}
\tag{34}
$$

We evaluate the remaining term:

$$
\begin{aligned}
\left\langle \hat{a}_I(t-\tau)\hat{a}_I^{(-)}(t) \right\rangle &= \left\langle g \right| \hat{a}_I^{(+)}(t-\tau)\hat{a}_I^{(-)}(t) \left| g \right\rangle \\
&= \sum_k d_{gk} d_{kg} e^{i\omega_{kg}\tau} \ ,
\end{aligned}
\tag{35}
$$

where we introduced the notation for the difference frequencies:

$$
\omega_{kj} = \omega_k - \omega_j \ .
\tag{36}
$$

The form of the field expectation value $\left\langle \hat{E}_I(t-\tau) \right\rangle$ depends, of course, on the state of the field. For a coherent state with known phase, it is given by Eq.(15) as:

$$
\begin{aligned}
\left\langle \hat{E}_I(t) \right\rangle &= \left\langle \alpha \right| \hat{E}^{(-)}(t) + \hat{E}^{(+)}(t) \left| \alpha \right\rangle \\
&= L_0 A^*(t) e^{i\omega_0 t} + L_0 A(t) e^{-i\omega_0 t} \ .
\end{aligned}
\tag{37}
$$

For a *single-photon state* the field expectation value is zero, as seen from Eqs.(15) and (20):

$$\langle \hat{E}_I(t) \rangle = \langle \varphi | \hat{E}^{(-)}(t) + \hat{E}^{(+)}(t) | \varphi \rangle = 0 \ . \tag{38}$$

The mean value of the field is zero because there is an unbalanced number of raising and lowering operators in the expectation value, resulting in projecting the vacuum state onto a single-photon state. This result can also be interpreted as saying that a single-photon state has no definite phase. Thus, in this case $\langle \hat{d}_I(t) \rangle = 0$.

But, even though the mean dipole is zero, the single-photon wave packet creates quantum state entanglement between the molecule and field, as can be seen by the following. The density operator to first order is $\hat{\rho}_I(t) = \hat{\rho}_0 + \hat{\rho}_I^{(1)}(t)$, where the initial state is $\hat{\rho}_0 = |g\rangle\langle g| \otimes |\varphi\rangle\langle\varphi|$, with $|\varphi\rangle$ the initial field state. Then from Eq.(31), after inserting the field operator and using the RWA:

$$\hat{\rho}_I^{(1)}(t) = \frac{i}{\hbar} L_0 \, d_{jg} |j\rangle\langle g| \int d\omega \, \varphi^*(\omega) e^{i\omega_{jg}t} e^{-i\omega t} \int_0^\infty d\tau e^{-\zeta\tau} \, e^{-i\omega_{jg}\tau} e^{i\omega\tau} |vac\rangle\langle\varphi| + \ hc \ . \tag{39}$$

where we assumed near resonance to a particular *g-j* transition. (To ensure convergence we inserted a small damping constant $\zeta$ in the exponential and set it to zero at the end). We carry out the integral and write the result as $\hat{\rho}_I^{(1)}(t) = \eta |j\rangle|vac\rangle\langle g|\langle\varphi| + \ hc$, where:

$$\eta = \frac{i}{\hbar} L_0 d_{jg} e^{i\omega_{jg}t} \int d\omega \, e^{-i\omega t} \frac{\varphi^*(\omega)}{i\omega_{jg} - i\omega} \ . \tag{40}$$

We can rewrite the result approximately as:

$$\begin{aligned} \hat{\rho}_I(t) &= |g\rangle|\varphi\rangle\langle g|\langle\varphi| + \eta |j\rangle|vac\rangle\langle g|\langle\varphi| + \eta^* |g\rangle|\varphi\rangle\langle j|\langle vac| + O(\eta^2) \\ &\simeq \big(|g\rangle|\varphi\rangle + \eta |j\rangle|vac\rangle\big)\big(\langle g|\langle\varphi| + \eta^* \langle j|\langle vac|\big) + O(\eta^2) \end{aligned}, \tag{41}$$

which we see represents a pure state $|g\rangle|\varphi\rangle + \eta |j\rangle|vac\rangle$.

The resulting state is entangled, that is it cannot be written in product form, $|\psi\rangle_M \otimes |\varphi\rangle_F$. Such entanglement provides the basic resource for many quantum-information techniques.

Note also that if two molecules are exposed to the same single-photon pulse and the photon is later observed to have been absorbed (it fails to trigger a 'perfect' detector), then the two molecules are left in an entangled state: they share the excitation of one photon.

Such entanglement is the basis for atomic-ensemble quantum memories, which store states of light coherently in an extended 'phased-array' of atoms. [45] Such an entangled state of many atoms can be thought of an exciton even if the atoms are distant from one another. If the electron spin is part of the state labels, this state is often called a spin wave.

As a second example, if the molecule is driven by a *coherent state*, the polarization density is:

$$\left\langle \hat{d}_I^{(-)}(t) \right\rangle = \frac{1}{i\hbar} \int_0^\infty d\tau \left\langle \hat{d}_I(t-\tau)\hat{d}_I^{(-)}(t) \right\rangle \left\langle \hat{E}_I(t-\tau) \right\rangle$$
$$\simeq \frac{L_0}{i\hbar} \sum_k d_{gk} d_{kg} \int_0^\infty d\tau e^{i\omega_{kg}\tau} \left( A^*(t-\tau) e^{i\omega_0(t-\tau)} + A(t-\tau) e^{-i\omega_0(t-\tau)} \right)$$

(42)

To simplify writing the result we will make the rotating-wave approximation (RWA), in which only near-resonant term $A^*(t-\tau)$ is kept. The non-resonant (counter rotating) terms do contribute as correction, which is often, but not always, small. we can express the result as:

$$\left\langle \hat{d}_I^{(-)}(t) \right\rangle = \frac{L_0}{i\hbar} e^{i\omega_0 t} \sum_k d_{gk} d_{kg} \int_0^\infty d\tau e^{i(\omega_{kg}-\omega_0)\tau} A^*(t-\tau)$$
$$\simeq \frac{L_0}{i\hbar} \sum_k d_{gk} d_{kg} e^{i\omega_{kg}t} \int_{-\infty}^t dt_1 e^{-i(\omega_{kg}-\omega_0)t_1} A^*(t_1)$$

(43)

where we transformed back to $t_1 = t - \tau$. We observe several facts from this result. If we evaluate the result at very long times, after the pulse has come and gone, then each term in the sum is proportional to the Fourier transform of $A(t)$, that is the spectral amplitude of the pulse, evaluated at the corresponding molecular transition frequency. In this case of coherent 'impulsive excitation,' (and ignoring damping and dephasing interactions) the polarization density oscillates at all of the excited molecular frequencies $\omega_{kg}$.

We leave as an exercise to show that after a coherent-state pulse interacts with the molecule, there is no entanglement between molecule and field. That is because when a pure coherent state experiences loss or absorption it is transformed into a pure coherent state of lesser amplitude. By definition, a pure state is not entangled with any other system.

### 5.2. Kubo theory of molecular dephasing

Now we consider the effects of dephasing interactions of the molecule with its surrounding environment, which may be a gas, liquid, or solid. To treat environmental perturbations rigorously one should introduce new terms into the Hamiltonian in Eq.(6). It turns out that a much simpler approach yields a realistic model—the Kubo theory of line shapes, in which the environment is treated as a (semiclassical) random process that causes the molecular energies to fluctuate randomly leading to finite transition line widths. For review see [30, 32, 46, 47, 48] To avoid a lengthy treatment, we illustrate only the simplest case of Kubo theory, that with zero-duration impact-like interactions of the molecule and the environment, leading to Lorentzian homogeneous line shapes.

In this phenomenological treatment, the molecular energies $\omega_{jk}$ are replaced by $\omega_{jk} + x_{jk}(t)$, where $x_{jk}(t)$ are random fluctuating quantities (frequencies) with zero mean. The factors of the form $\exp(-i\omega t)$ (dropping the $jk$ subscript) that appear in equations such as Eq.(42) are replaced by factors of the form:

$$\left\langle \exp\left(-i\int_0^t dt'[\omega + x(t')]\right)\right\rangle = \exp(-i\omega t)\left\langle \exp(-i\phi(t))\right\rangle , \qquad (44)$$

where the random accumulated phase is:

$$\phi(t) = \int_0^t dt' x(t') . \qquad (45)$$

Here the brackets indicate an ensemble average (not a time average) over possible realizations of the random process $x(t)$. It is assumed that the random process $x(t)$ (and thus also $\phi(t)$) obeys Gaussian statistics, namely the probability for $\phi(t)$ to take on a particular value is given by:

$$Pr(\phi) = \frac{1}{\sqrt{2\pi\sigma^2}} e^{-\phi^2/2\sigma^2} , \qquad (46)$$

where the variance of $\phi$ is given by:

$$\begin{aligned}
\sigma^2 &= \left\langle \phi^2(t)\right\rangle \\
&= \int_0^t dt' \int_0^t dt'' \left\langle x(t')x(t'')\right\rangle .
\end{aligned} \qquad (47)$$

Thus, the average in Eq.(44) becomes:

$$\begin{aligned}
\left\langle \exp(-i\phi(t))\right\rangle &= \int_{-\infty}^{\infty} d\phi\, Pr(\phi)\exp(-i\phi(t)) \\
&= \exp[-\sigma^2/2] .
\end{aligned} \qquad (48)$$

In the case of zero-duration impact-like interactions of the molecule and the environment, the correlation function is delta-like, that is $\left\langle x(t)x(t')\right\rangle = \hbar^2 2\gamma\,\delta(t-t')$, where $2\gamma$ is the linewidth of the considered transition, as we will see. The integral is easily carried out:

$$\begin{aligned}
\sigma^2 &= \int_0^t dt' \int_0^t dt'' 2\gamma\,\delta(t'-t'') \\
&= 2\gamma t ,
\end{aligned} \qquad (49)$$

thereby giving:

$$\left\langle \exp(-i\phi(t))\right\rangle = \exp[-\gamma t] . \qquad (50)$$

Recall that $t$ is positive. We then have, replacing Eq.(42):

$$\left\langle \hat{d}_I^{(-)}(t) \right\rangle = \frac{L_0}{i\hbar} e^{i\omega_0 t} \sum_k d_{gk} d_{kg} \int_0^\infty d\tau \, e^{-[\gamma_{kg} - i(\omega_{kg} - \omega_0)]\tau} A^*(t-\tau) \; . \tag{51}$$

The simple version of Kubo dephasing theory does not include the possibility of population damping—decay of the probability for the molecule to be in a particular state resulting from spontaneous emission or other incoherent nonradiative processes that couple two states. While such processes can be treated rigorously, [e.g, 46, 47] here we treat them phenomenologically by adding additional damping rates into the Kubo dephasing rates. The rules are: if the population decay rate out of each state $i$ is denoted $A_i$, then the population $\rho_{ii}$ is damped as $\exp[-A_i t]$ and the off-diagonal density matrix element $\rho_{ij}$ of the transition $i \leftrightarrow j$ is damped as $\exp[-(A_i/2 + A_j/2)t]$. [e.g, 46, Sec. 14.5] Population damping causes a broadening of the spectral lines, called lifetime broadening. There can also be population-increasing processes, such as spontaneous emission into a state, but for our purposes these processes won't be important because we treat the problem perturbatively and typically on short time scales compared to the times scales of such effects.

### 5.3. Steady-state induced dipole

Now that we have dephasing included, we can evaluate the polarization density in steady state, in which the field amplitude $A$ is constant:

$$\begin{aligned}
\left\langle \hat{d}_I^{(-)}(t) \right\rangle &= \frac{L_0}{i\hbar} e^{i\omega_0 t} \sum_k d_{gk} d_{kg} \int_0^\infty d\tau \, e^{-[\gamma_{kg} - i(\varepsilon_{kg} - \omega_0)]\tau} A^*(t-\tau) \\
&= \frac{L_0}{i\hbar} A^* e^{i\omega_0 t} \sum_k d_{gk} d_{kg} \int_0^\infty d\tau \, e^{-[\gamma_{kg} - i(\varepsilon_{kg} - \omega_0)]\tau} \\
&= \frac{L_0}{i\hbar} A^* e^{i\omega_0 t} \sum_k d_{gk} d_{kg} \frac{1}{\gamma_{kg} - i(\varepsilon_{kg} - \omega_0)} \; .
\end{aligned} \tag{52}$$

We see that in the steady-state regime the polarization oscillates at the driving frequency $\omega_0$, whereas we saw earlier that in the impulsive regime it oscillates at the molecular frequencies. These results are, of course, the same as obtained in a semiclassical treatment where the field is treated as a classical one.

### 5.4. One-photon-induced population

#### 5.4.1 General case

The population $P_e^{(2)}$ in a given excited state $|e\rangle$ resulting from one-photon absorption equals the expectation value $P_e^{(2)} = Tr\left( \hat{\rho}_I^{(2)}(t) |e\rangle \langle e| \right)$, where $\hat{\rho}_I^{(2)}(t)$ is the density operator solution iterated to second order:

$$\hat{\rho}_I{}^{(2)}(t) = \left(\frac{1}{i\hbar}\right)^2 \int\limits_{-\infty}^{t} dt_2 \int\limits_{-\infty}^{t_2} dt_1 \, C^{(2)} \quad,$$

$$C^{(2)} = \left[\hat{H}_I(t_2), \left[\hat{H}_I(t_1), \hat{\rho}_0\right]\right] \tag{53}$$

$$= \hat{H}_I(t_2)\hat{H}_I(t_1)\hat{\rho}_0 - \hat{H}_I(t_2)\hat{\rho}_0\hat{H}_I(t_1) - \hat{H}_I(t_1)\hat{\rho}_0\hat{H}_I(t_2) + \hat{\rho}_0\hat{H}_I(t_1)\hat{H}_I(t_2).$$

Because the initial state is $\hat{\rho}_M = |g\rangle\langle g|$, we note that $\hat{\rho}_0|e\rangle\langle e| = |e\rangle\langle e|\hat{\rho}_0 = 0$ and thus only two of the four terms contribute to the trace:

$$C^{(2)} \rightarrow -\hat{H}_I(t_2)\hat{\rho}_0\hat{H}_I(t_1) - \hat{H}_I(t_1)\hat{\rho}_0\hat{H}_I(t_2) \quad. \tag{54}$$

For near-resonant excitation, we again apply the RWA, retaining only terms in which the molecule and field exchange excitations:

$$\hat{H}_I(t) = -\hat{d}_I(t)^{(-)}\hat{E}_I{}^{(+)}(t) - \hat{d}_I(t)^{(+)}\hat{E}_I{}^{(-)}(t) \quad. \tag{55}$$

The neglected terms will contribute small corrections to the final result, which are significant only when the excitation is far from resonance. Using Eq.(9) for the dipole operators, we find for the first term in Eq.(54):

$$\hat{H}_I(t_2)\hat{\rho}_M\hat{H}_I(t_1) =$$

$$= \hat{d}_I{}^{(-)}(t_2)\hat{E}_I{}^{(+)}(t_2)|g\rangle\langle g|\hat{d}_I{}^{(+)}(t_1)\hat{E}_I{}^{(-)}(t_1) \tag{56}$$

$$= \sum_{i,j} d_{ig}d_{gj}|i\rangle\langle j| e^{i\omega_{ig}t_2}e^{i\omega_{gj}t_1}\hat{E}_I{}^{(+)}(t_2)\hat{E}_I{}^{(-)}(t_1).$$

where $\omega_{ij} = \omega_i - \omega_j$. The second term in Eq.(54) is the same with $t_1$ and $t_2$ swapped. So, using permutation inside the trace and $Tr(\hat{A}^\dagger) = Tr(\hat{A})^*$, and the initial state $\hat{\rho}_0 = |g\rangle\langle g| \otimes \hat{\rho}_F$, we have for the population, or probability to find the molecule in state $e$:

$$P_e^{(2)} = \frac{1}{\hbar^2}\int\limits_{-\infty}^{t} dt_2 \int\limits_{-\infty}^{t_2} dt_1 \, Tr\left(\hat{H}_I(t_2)\hat{\rho}_0\hat{H}_I(t_1)|e\rangle\langle e|\right) + cc$$

$$= \frac{1}{\hbar^2}\int\limits_{-\infty}^{t} dt_2 \int\limits_{-\infty}^{t_2} dt_1 \, Tr_M\left(\sum_i d_{ig}d_{ge}|i\rangle\langle e| e^{i\omega_{ig}t_2}e^{i\omega_{ge}t_1}\right)Tr_F\left(\hat{E}_I{}^{(+)}(t_2)\hat{\rho}_F\hat{E}_I{}^{(-)}(t_1)\right) + cc \tag{57}$$

$$= \frac{1}{\hbar^2}d_{ge}d_{eg}\int\limits_{-\infty}^{\infty} dt_2 \int\limits_{-\infty}^{t_2} dt_1 \, e^{i\omega_{eg}(t_2-t_1)}\left\langle \hat{E}_I{}^{(-)}(t_1)\hat{E}_I{}^{(+)}(t_2)\right\rangle + cc \quad.$$

where the second-order field correlation function is:

$$\left\langle \hat{E}_I{}^{(-)}(t_1)\hat{E}_I{}^{(+)}(t_2)\right\rangle = Tr_F\left(\hat{\rho}_F\hat{E}_I{}^{(-)}(t_1)\hat{E}_I{}^{(+)}(t_2)\right) \quad. \tag{58}$$

In the final line we took the long-time limit, $t = \infty$, to find the population immediately after the pulse has passed through the molecule. The final population is equivalent to the pathway represented in **Fig. 1b** and **1c**. The second-order field correlation function, when divided by $\left\langle \hat{E}_I^{(-)}(t_1)\hat{E}_I^{(+)}(t_1) \right\rangle^{1/2} \left\langle \hat{E}_I^{(-)}(t_2)\hat{E}_I^{(+)}(t_2) \right\rangle^{1/2}$, is often referred to as the degree of first-order temporal coherence $g^{(1)}(t_1, t_2)$ [28]

Now, we insert the dephasing factor according to Kubo theory, we transform to a difference-time variable, $\tau = t_2 - t_1$, and rename the remaining integration variable $t_2 \equiv t$ Then the solution is:

$$P_e^{(2)} = \frac{1}{\hbar^2} d_{ge} d_{eg} \int_{-\infty}^{\infty} dt \int_0^{\infty} d\tau \, e^{-(\gamma_{eg} - i\omega_{eg})\tau} \left\langle \hat{E}_I^{(-)}(t-\tau)\hat{E}_I^{(+)}(t) \right\rangle + cc \ . \tag{59}$$

It is useful to transform to the frequency domain by inserting Eq.(15) twice:

$$\begin{aligned}
P_e^{(2)} &= \frac{1}{\hbar^2} L_0^{\ 2} d_{ge} d_{eg} \int d\omega \int_0^{\infty} d\tau \, e^{-(\gamma_{eg} - i\omega_{eg} + i\omega)\tau} \left\langle \hat{a}^{\dagger}(\omega)\hat{a}(\omega) \right\rangle + cc \\
&= \frac{1}{\hbar^2} L_0^{\ 2} d_{ge} d_{eg} \int d\omega \, \frac{\left\langle \hat{a}^{\dagger}(\omega)\hat{a}(\omega) \right\rangle}{\gamma_{eg} - i(\omega_{eg} - \omega)} \ + \ cc \\
&= \frac{1}{\hbar^2} L_0^{\ 2} d_{ge} d_{eg} \int d\omega \, \frac{2\gamma_{eg}}{\gamma_{eg}^{\ 2} + (\omega_{eg} - \omega)^2} \left\langle \hat{a}^{\dagger}(\omega)\hat{a}(\omega) \right\rangle \ .
\end{aligned} \tag{60}$$

This result has a simple interpretation—the photons at frequency $\omega$, whose number is $\left\langle \hat{a}^{\dagger}(\omega)\hat{a}(\omega) \right\rangle$, drive the transition with strength given by the Lorentzian line shape.

### 5.4.2 One-photon absorption: Coherent state

With these results in hand, we can evaluate the one-photon excitation probability for a coherent-state pulse, for which $\left\langle \hat{E}_I^{(-)}(t_2)\hat{E}_I^{(+)}(t_1) \right\rangle = L_0^{\ 2} A^*(t_2)A(t_1)e^{i\omega_0(t_2 - t_1)}$, we find:

$$P_e^{(2)} = \frac{1}{\hbar^2} L_0^{\ 2} d_{ge} d_{eg} \int_{-\infty}^{\infty} dt \int_0^{\infty} d\tau \, e^{-(\gamma_{eg} - i\omega_{eg} + i\omega_0)\tau} A^*(t-\tau)A(t) + cc \ . \tag{61}$$

Or, in the frequency domain, for which $\left\langle \hat{a}^{\dagger}(\omega)\hat{a}(\omega) \right\rangle = \alpha^*(\omega)\alpha(\omega)$, we have $\left\langle \hat{a}^{\dagger}(\omega)\hat{a}(\omega) \right\rangle = |\alpha_0|^2 \varphi^*(\omega)\varphi(\omega)$ and:

$$P_e^{(2)} = \frac{1}{\hbar^2} L_0^{\ 2} d_{ge} d_{eg} \int d\omega \, \frac{2\gamma_{eg}}{\gamma_{eg}^{\ 2} + (\omega_{eg} - \omega)^2} |\alpha_0|^2 \varphi^*(\omega)\varphi(\omega) \ . \tag{62}$$

For quasi-monochromatic light, we can write the result in terms of the absorption cross section $\sigma^{(1)}$. The probability to excite the molecule per photon incident should equal the ratio of $\sigma^{(1)}$ to the optical beam's cross-sectional area $A_0$ Recall the mean number of photons is $N = \int |\alpha(\omega)|^2 \, d\omega$. Then, if the spectrum of light is centered at $\omega_0$ and narrow compared to the Lorentzian absorption line, we have (using the definition of $L_0$):

$$P_e^{(2)} = N \frac{\sigma^{(1)}}{A_0} \ , \tag{63}$$

where the one-photon cross section (with units $m^{-2}$) is:

$$\sigma^{(1)} = \frac{\omega_0}{\varepsilon_0 \hbar n c} \left| d_{eg} \right|^2 \frac{\gamma_{eg}}{\gamma_{eg}^2 + (\omega_{eg} - \omega_0)^2} \ . \tag{64}$$

This result agrees with, for example, [49].

### 5.4.3 One-photon absorption: Single-photon state

Next, we evaluate the one-photon excitation probability when driven by a single-photon state, Eq.(20). First, we note that operating the annihilation operator on the state gives:

$$\begin{aligned}
\hat{a}(\omega)|\varphi\rangle &= \int d\omega' \varphi^*(\omega') \hat{a}(\omega) \hat{a}^\dagger(\omega') |vac\rangle \\
&= \int d\omega' \varphi^*(\omega') \left\{ 2\pi \delta(\omega' - \omega) + \hat{a}^\dagger(\omega') \hat{a}(\omega) \right\} |vac\rangle \\
&= \varphi^*(\omega) |vac\rangle .
\end{aligned} \tag{65}$$

Thus, we have $\left\langle \hat{a}^\dagger(\omega) \hat{a}(\omega) \right\rangle = \varphi^*(\omega) \varphi(\omega)$, and Eq.(59) gives:

$$P_e^{(2)} = \frac{1}{\hbar^2} L_0^2 d_{ge} d_{eg} \int d\omega \frac{2\gamma_{eg}}{\gamma_{eg}^2 + (\omega_{eg} - \omega)^2} \varphi^*(\omega) \varphi(\omega) \ . \tag{66}$$

We see that the probabilities to excite the state $e$ by a coherent state or by a single-photon state have precisely the same form, except that for the former we have $\int d\omega |\alpha(\omega)|^2 = N$ and in the latter we have $\int d\omega |\varphi(\omega)|^2 = 1$. Therefore, when exciting a single molecule in the linear-response regime, a single-photon pulse has the same effect as a coherent-state pulse with mean photon number equal to one. In this scenario there is nothing especially 'quantum' about single-photon absorption. As pointed out earlier, the story is different when exciting two or more molecules—the single-photon pulse can create entanglement of excitation in the molecules' joint state, whereas a coherent-state pulse does not.

## 6. Two-photon absorption: General treatment

Here we give the results of the fourth-order solution of the density matrix, setting up its application to TPA by coherent states of EPP. And we discuss the 'quantum advantages' of EPP relative to coherent states.

### 6.1. Conventional TPA cross section

Before giving general results, we touch base with the well-known lore on the two-photon cross section, first calculated in 1931 by Maria Göppert-Mayer, named a Nobel Laureate in Physics in 1963. Of special interest is the role, if any, of the NRP and RP pathways in far-off-resonance TPA. In the conventional theory these terms do not appear, and the question is why? The conventional theory uses second-order perturbation theory for state amplitudes, rather than density-matrix elements, assuming the intermediate states are far from resonance with the exciting field's frequency, and so dephasing rates are ignored (as they must be in that treatment). See Boyd, Sec. 12.5.3, for a concise review. [29] The result is expressed in terms of a TPA cross section:

$$\sigma^{(2)} = \left( \frac{\omega_0}{\hbar \varepsilon_0 n c} \right)^2 \frac{\pi}{2\pi \gamma_{fg}} \left| \sum_e \frac{d_{ef} d_{ge}}{\omega_{eg} - \omega_0} \right|^2 , \qquad (67)$$

where $1/2\pi \gamma_{fg}$ plays the role of the density of states for the transition. We will see how to reproduce this result using the density matrix approach, which provides further insight into the role of the different pathways.

### 6.2. Fourth-order TPA solution

In the density matrix approach the fourth-order solution, needed to calculate the probability for TPA, is:

$$\hat{\rho}^{(4)}(t) = \sum_{p,q,r,s} \int_{-\infty}^{t} dt_4 \int_{-\infty}^{t_4} dt_3 \int_{-\infty}^{t_3} dt_2 \int_{-\infty}^{t_2} dt_1 \left[ \hat{V}^{(s)}(t_4), \left[ \hat{V}^{(r)}(t_3), \left[ \hat{V}^{(q)}(t_2), \left[ \hat{V}^{(p)}(t_1), \hat{\rho}_0 \right] \right] \right] \right], \qquad (68)$$

where the sum is over $p,q,r,s = \pm$, and, for simplicity, we made the RWA and used the compact notation:

$$\hat{V}^{(p)}(t) = \hat{\mu}^{(-p)}(t) \hat{E}^{(p)}(t) \quad , \quad p = \pm , \qquad (69)$$

where $\hat{\mu}^{(-)}(t) = \hat{d}^{(-)}(t)/\hbar$ is a scaled dipole operator. $\hat{V}^{(+)}$ raises the molecule and lowers the field. $\hat{V}^{(-)}$ does the opposite. The solution has sixteen terms contributing to the sum, many of which can be neglected in most cases. We focus on those TPA terms that lead to population in the $f$ state, $P_f = Tr \, \hat{\rho}^{(4)} |f\rangle\langle f|$. Here and in the following we drop the subscript $I$ indicating the Interaction Picture.

Given that the transition of interest is $g \to f$, the only terms resulting from the nested commutators that are nonzero are those of the form $V^{(+)}V^{(+)}\hat{\rho}_0 V^{(-)}V^{(-)}$, giving:

$$P_f = \int\limits_{-\infty}^{t} dt_4 \int\limits_{-\infty}^{t_4} dt_3 \int\limits_{-\infty}^{t_3} dt_2 \int\limits_{-\infty}^{t_2} dt_1 \left(Q_{DQC} + Q_{NRP} + Q_{RP}\right) + cc, \tag{70}$$

where we defined:

$$\begin{aligned}
Q_{DQC} &= Tr_M Tr_F \left(\hat{V}^{(+)}(t_4)\hat{V}^{(+)}(t_3)\hat{\rho}_0 \hat{V}^{(-)}(t_1)\hat{V}^{(-)}(t_2)\big|f\rangle\langle f\big|\right) \\
Q_{NRP} &= Tr_M Tr_F \left(\hat{V}^{(+)}(t_4)\hat{V}^{(+)}(t_2)\hat{\rho}_0 \hat{V}^{(-)}(t_1)\hat{V}^{(-)}(t_3)\big|f\rangle\langle f\big|\right) \\
Q_{RP} &= Tr_M Tr_F \left(\hat{V}^{(+)}(t_3)\hat{V}^{(+)}(t_2)\hat{\rho}_0 \hat{V}^{(-)}(t_1)\hat{V}^{(-)}(t_4)\big|f\rangle\langle f\big|\right).
\end{aligned} \tag{71}$$

To interpret these terms, consider that all operators act toward $\hat{\rho}_0$, in the order $t_1, t_2, t_3, t_4$. Referring to **Fig. 2**, we can see clearly the correspondence of each term with each diagram.

Because we are using the Interaction Picture, operators for the field commute with those of the molecule. Thus, we can separate the correlation functions as follows, keeping in mind that $\hat{\rho}_M = \big|g\rangle\langle g\big|$:

$$\begin{aligned}
Q_{DQC} &= C_M^{DQC} C_F^{DQC} \\
Q_{NRP} &= C_M^{NRP} C_F^{NRP} \\
Q_{RP} &= C_M^{RP} C_F^{RP}.
\end{aligned} \tag{72}$$

where the molecule correlation functions are:

$$\begin{aligned}
C_M^{DQC} &= Tr_M \left(\hat{\mu}^{(-)}(t_4)\hat{\mu}^{(-)}(t_3)\hat{\rho}_M \hat{\mu}^{(+)}(t_1)\hat{\mu}^{(+)}(t_2)\big|f\rangle\langle f\big|\right) \\
C_M^{NRP} &= Tr_M \left(\hat{\mu}^{(-)}(t_4)\hat{\mu}^{(-)}(t_2)\hat{\rho}_M \hat{\mu}^{(+)}(t_1)\hat{\mu}^{(+)}(t_3)\big|f\rangle\langle f\big|\right) \\
C_M^{RP} &= Tr_M \left(\hat{\mu}^{(-)}(t_3)\hat{\mu}^{(-)}(t_2)\hat{\rho}_M \hat{\mu}^{(+)}(t_1)\hat{\mu}^{(+)}(t_4)\big|f\rangle\langle f\big|\right).
\end{aligned} \tag{73}$$

These are the same forms that appear in the semiclassical treatment, where the field is treated as a classical function. [30, 32, 46] They are evaluated in **Appendix A**, which derives the following results using Kubo dephasing theory:

$$\begin{aligned}
C_M^{DQC} &= \sum_{e,e'} \mu_{fe'}\mu_{e'g}\mu_{ge}\mu_{ef} \, e^{-(\gamma_{fe}-i\omega_{fe})r} e^{-(\gamma_{fg}-i\omega_{fg})s} e^{-(\gamma_{eg}-i\omega_{eg})\tau} \\
C_M^{NRP} &= \sum_{e,e'} \mu_{fe'}\mu_{e'g}\mu_{ge}\mu_{ef} \, e^{-(\gamma_{fe}-i\omega_{fe})r} e^{-(\gamma_{ee'}-i\omega_{ee'})s} e^{-(\gamma_{eg}-i\omega_{eg})\tau} \\
C_M^{RP} &= \sum_{e,e'} \mu_{fe'}\mu_{e'g}\mu_{ge}\mu_{ef} \, e^{-(\gamma_{fe}+i\omega_{fe})r} e^{-(\gamma_{ee'}-i\omega_{ee'})s} e^{-(\gamma_{eg}-i\omega_{eg})\tau},
\end{aligned} \tag{74}$$

where we introduced the difference-time variables $r = t_4 - t_3$, $s = t_3 - t_2$, $\tau = t_2 - t_1$. These variables track the time increases during disjoint time intervals during which dephasing takes place, allowing the dephasing during each interval to be considered separately as in Eq.(74). For each transition at frequency $\omega_{ij}$ the corresponding dephasing rate is $\gamma_{ij}$. Note the plus sign in the exponential argument $(\gamma_{fe} + i\omega_{fe})r$ of the RP correlation function, which is related to the well-known effect of photon echoes in special cases. [31, 30]

The field correlation functions in Eq.(72), which can describe quantum states of light, are given by:

$$C_F^{DQC} = Tr_F\left(\hat{E}^{(+)}(t_4)\hat{E}^{(+)}(t_3)\hat{\rho}_F\hat{E}^{(-)}(t_1)\hat{E}^{(-)}(t_2)\right)$$

$$C_F^{NRP} = Tr_F\left(\hat{E}^{(+)}(t_4)\hat{E}^{(+)}(t_2)\hat{\rho}_F\hat{E}^{(-)}(t_1)\hat{E}^{(-)}(t_3)\right) \qquad (75)$$

$$C_F^{RP} = Tr_F\left(\hat{E}^{(+)}(t_3)\hat{E}^{(+)}(t_2)\hat{\rho}_F\hat{E}^{(-)}(t_1)\hat{E}^{(-)}(t_4)\right).$$

We can summarize them using a common form, where the $t_a$, etc. are arbitrary ('dummy') variables:

$$\begin{aligned}C_F(t_a, t_b, t_c, t_d) &= Tr_F\left(\hat{E}^{(+)}(t_a)\hat{E}^{(+)}(t_b)\hat{\rho}_F\hat{E}^{(-)}(t_c)\hat{E}^{(-)}(t_d)\right) \\ &= Tr_F\left(\hat{\rho}_F\hat{E}^{(-)}(t_c)\hat{E}^{(-)}(t_d)\hat{E}^{(+)}(t_a)\hat{E}^{(+)}(t_b)\right) \\ &= \left\langle\hat{E}^{(-)}(t_c)\hat{E}^{(-)}(t_d)\hat{E}^{(+)}(t_a)\hat{E}^{(+)}(t_b)\right\rangle,\end{aligned} \qquad (76)$$

where we used cyclic permutation inside the trace. This fourth-order field correlation function, when divided by $\left\langle\hat{E}_I^{(-)}(t_1)\hat{E}_I^{(+)}(t_1)\right\rangle\left\langle\hat{E}_I^{(-)}(t_2)\hat{E}_I^{(+)}(t_2)\right\rangle$, is often referred to as the degree of second-order temporal coherence $g^{(2)}(t_c, t_d, t_a, t_b)$. [28]

For pure states it can be written:

$$C_F(t_a, t_b, t_c, t_d) = \left\langle\Psi\right|\hat{E}^{(-)}(t_c)\hat{E}^{(-)}(t_d)\hat{E}^{(+)}(t_a)\hat{E}^{(+)}(t_b)\left|\Psi\right\rangle. \qquad (77)$$

In the case of a pure two-photon state $\left|\Psi^{(2)}\right\rangle$ the correlation function can be written:

$$C_F^{(2)} = \left\langle\Psi^{(2)}\right|\hat{E}^{(+)}(t_c)\hat{E}^{(-)}(t_d)\left|vac\right\rangle\left\langle vac\right|\hat{E}^{(+)}(t_a)\hat{E}^{(+)}(t_b)\left|\Psi^{(2)}\right\rangle, \qquad (78)$$

where we inserted unity, $\sum_\Psi\left|\Psi\right\rangle\left\langle\Psi\right|$, in the center of Eq.(77), and noted that only the vacuum term contributes. It is notable that in this case the correlation function equals the product of two functions, known as the two-photon detection amplitude:

$$\Phi(t_a, t_b) = \left\langle vac\right|\hat{E}^{(+)}(t_a)\hat{E}^{(+)}(t_b)\left|\Psi^{(2)}\right\rangle. \qquad (79)$$

This result suggests we can view TPA in a molecule as a two-photon detector. [9]

### 6.3. Two-photon amplitude

The form of the two-photon detection amplitude reveals interesting aspects of quantum optics, namely the role played by the boson nature of photons when viewed (with due circumspect) as particles. We insert the two-photon component of the state in Eq.(28) into the expression Eq.(78), and insert the frequency representation of the field operators from Eq.(15):

$$
\begin{aligned}
\Phi(t_a, t_b) &= \left\langle vac \left| \hat{E}^{(+)}(t_a) \hat{E}^{(+)}(t_b) \varepsilon \int d\omega \int d\tilde{\omega} \, \psi(\omega, \tilde{\omega}) \hat{a}^\dagger(\omega) \hat{a}^\dagger(\tilde{\omega}) \right| vac \right\rangle \\
&= \varepsilon L_0^2 \int d\omega \int d\tilde{\omega} \int d\omega' \int d\tilde{\omega}' \, \psi(\omega, \tilde{\omega}) e^{-i\omega' t_a} e^{-i\tilde{\omega}' t_b} C_{\omega, vac}^{(4)} \ .
\end{aligned}
\tag{80}
$$

where the frequency-domain correlation function in the vacuum state is:

$$
C_{\omega, vac}^{(4)} = \left\langle vac \left| \hat{a}(\omega') \hat{a}(\tilde{\omega}') \hat{a}^\dagger(\omega) \hat{a}^\dagger(\tilde{\omega}) \right| vac \right\rangle
\tag{81}
$$

Using the commutator $[\hat{a}_j(\omega), \hat{a}_i^\dagger(\omega')] = 2\pi\delta(\omega - \omega')\delta_{ij}$, we find:

$$
\begin{aligned}
C_{\omega, vac}^{(4)} &= \left\langle vac \left| \hat{a}(\omega') \hat{a}(\tilde{\omega}') \hat{a}^\dagger(\omega) \hat{a}^\dagger(\tilde{\omega}) \right| vac \right\rangle \\
&= (2\pi)^2 \delta(\omega' - \omega)\delta(\tilde{\omega}' - \tilde{\omega}) + (2\pi)^2 \delta(\omega' - \tilde{\omega})\delta(\tilde{\omega}' - \omega).
\end{aligned}
\tag{82}
$$

Then Eq.(79) becomes:

$$
\begin{aligned}
\Phi(t_a, t_b) &= \varepsilon L_0^2 \int d\omega \int d\tilde{\omega} \, \psi(\omega, \tilde{\omega}) e^{-i\omega t_a} e^{-i\tilde{\omega} t_b} + \varepsilon L_0^2 \int d\omega \int d\tilde{\omega} \, \psi(\omega, \tilde{\omega}) e^{-i\tilde{\omega} t_a} e^{-i\omega t_b} \\
&= 2\varepsilon L_0^2 \int d\omega \int d\tilde{\omega} \left\{ \frac{\psi(\omega, \tilde{\omega}) + \psi(\tilde{\omega}, \omega)}{2} \right\} e^{-i\omega t_a} e^{-i\tilde{\omega} t_b} \ .
\end{aligned}
\tag{83}
$$

where in the last line we swapped the $\omega, \tilde{\omega}$ variables. We thus see the interesting result that the two-photon detection amplitude depends only on the symmetrized form of the JSA, which we denote as:

$$
\Psi(\omega, \tilde{\omega}) = \frac{\psi(\omega, \tilde{\omega}) + \psi(\tilde{\omega}, \omega)}{2} \ .
\tag{84}
$$

But recall we are treating, for simplicity, only the case that the driving light is in a single spatial-polarization mode of the field. After Eq.(28) we noted that in this case the JSA must be symmetric, that is, $\psi(\tilde{\omega}, \omega) = \psi(\omega, \tilde{\omega})$. Thus, in this case $\Psi(\omega, \tilde{\omega}) = \psi(\omega, \tilde{\omega})$. We can thus write:

$$
\Phi(t_a, t_b) = 2\varepsilon L_0^2 \int d\omega \int d\tilde{\omega} \, \Psi(\omega, \tilde{\omega}) e^{-i\omega t_a} e^{-i\tilde{\omega} t_b} \ .
\tag{85}
$$

The mod-square $|\Phi(t_a, t_b)|^2$ is the joint probability to detect a photon at time $t_a$ and a photon at time $t_b$, presuming one has sufficiently fast detectors. If one separates the field into its spectrum, say, using a prism, then $|\Psi(\omega,\tilde{\omega})|^2$ is seen to be the joint probability to detect a pair of photons at the two indicated frequencies.

We retain the notation $\Psi(\omega,\tilde{\omega})$ because it can be shown straightforwardly to arise automatically even in the case of Type-II SPDC, where the signal and idler modes as distinct, so the state is written with labels on the creation operators:

$$\left|\Psi^{(2)}\right\rangle = \varepsilon \int d\omega \int d\tilde{\omega}\, \psi(\omega,\tilde{\omega}) \hat{a}_I^\dagger(\omega) \hat{a}_S^\dagger(\tilde{\omega})\left|vac\right\rangle . \tag{86}$$

The JSA here need not be symmetric, but the symmetrized form still determines the two-photon detection amplitude as in Eq.(83). This fact illustrates an important point in quantum optics: modes of the field are distinct and therefore in the quantum sense they are distinguishable. Therefore, the joint state, $\psi(\omega,\tilde{\omega})$, of two modes need not be symmetric under label exchange. On the other hand, when viewing light as made of photons, which in the quantum sense are indistinguishable, their joint state $\Psi(\omega,\tilde{\omega})$ must by symmetric. The needed symmetry is automatically satisfied by the math of boson commutators.

### 6.4. TPA probabilities

Transforming to difference-time variables, $r = t_4 - t_3$, $s = t_3 - t_2$, $\tau = t_2 - t_1$, and the inverse, $t_1 = (t_4 - r - s - \tau)$, $t_2 = (t_4 - r - s)$, $t_3 = (t_4 - r)$, and taking $t$ to infinity to encompass the entire excitation pulse, we find from Eq.(70) and (74):

$$\begin{aligned}
P_f^{DQC} &= \sum_{e,e'} M_{fe'eg} R_{e,e'}^{DQC} + cc \\
P_f^{NRP} &= \sum_{e,e'} M_{fe'eg} R_{e,e'}^{NRP} + cc \\
P_f^{RP} &= \sum_{e,e'} M_{fe'eg} R_{e,e'}^{RP} + cc ,
\end{aligned} \tag{87}$$

where again we denote the product of matrix elements by $M_{fe'eg} = \mu_{fe'}\mu_{e'g}\mu_{ge}\mu_{ef}$, and:

$$R_{e,e'}^{DQC} = \int\limits_0^\infty dr \int\limits_0^\infty ds \int\limits_0^\infty d\tau\, e^{-(\gamma_{fe}-i\omega_{fe'})r}\, e^{-(\gamma_{fg}-i\omega_{fg})s}\, e^{-(\gamma_{eg}-i\omega_{eg})\tau} \left\langle \hat{E}^{(-)}(t_1)\hat{E}^{(-)}(t_2)\hat{E}^{(+)}(t_4)\hat{E}^{(+)}(t_3)\right\rangle$$

$$R_{e,e'}^{NRP} = \int\limits_0^\infty dr \int\limits_0^\infty ds \int\limits_0^\infty d\tau\, e^{-(\gamma_{fe}-i\omega_{fe'})r}\, e^{-(\gamma_{ee'}-i\omega_{ee'})s}\, e^{-(\gamma_{eg}-i\omega_{eg})\tau} \left\langle \hat{E}^{(-)}(t_1)\hat{E}^{(-)}(t_3)\hat{E}^{(+)}(t_4)\hat{E}^{(+)}(t_2)\right\rangle \tag{88}$$

$$R_{e,e'}^{RP} = \int\limits_0^\infty dr \int\limits_0^\infty ds \int\limits_0^\infty d\tau\, e^{-(\gamma_{fe}+i\omega_{fe'})r}\, e^{-(\gamma_{ee'}-i\omega_{ee'})s}\, e^{-(\gamma_{eg}-i\omega_{eg})\tau} \left\langle \hat{E}^{(-)}(t_1)\hat{E}^{(-)}(t_4)\hat{E}^{(+)}(t_3)\hat{E}^{(+)}(t_2)\right\rangle .$$

These expressions are equivalent to those in [30, 32, 46], generalized here to arbitrary states of the field.

Inserting the frequency-domain form of the field operators, Eq.(15), and after some laborious math, the time integrals can be evaluated to give:

$$R_{e,e'}^{DQC} = L_0^4 \int d\omega' \int d\omega \int d\tilde{\omega} \frac{\left\langle \hat{a}^\dagger(\omega')\hat{a}^\dagger(\tilde{\omega}')\hat{a}(\omega)\hat{a}(\tilde{\omega}) \right\rangle}{(\gamma_{fe'} - i\omega_{fe'} + i\tilde{\omega}')(\gamma_{fg} - i\omega_{fg} + i\omega + i\tilde{\omega})(\gamma_{eg} - i\omega_{eg} + i\omega')}$$

$$R_{e,e'}^{NRP} = L_0^4 \int d\omega' \int d\omega \int d\tilde{\omega} \frac{\left\langle \hat{a}^\dagger(\omega')\hat{a}^\dagger(\tilde{\omega}')\hat{a}(\omega)\hat{a}(\tilde{\omega}) \right\rangle}{(\gamma_{fe'} - i\omega_{fe'} + i\tilde{\omega}')(\gamma_{ee'} - i\omega_{ee'} + i\omega' - i\omega)(\gamma_{eg} - i\omega_{eg} + i\omega')} \quad (89)$$

$$R_{e,e'}^{RP} = L_0^4 \int d\omega' \int d\omega \int d\tilde{\omega} \frac{\left\langle \hat{a}^\dagger(\omega')\hat{a}^\dagger(\tilde{\omega}')\hat{a}(\omega)\hat{a}(\tilde{\omega}) \right\rangle}{(\gamma_{fe} + i\omega_{fe} - i\tilde{\omega}')(\gamma_{ee'} - i\omega_{ee'} + i\omega' - i\omega)(\gamma_{eg} - i\omega_{eg} + i\omega')},$$

where for compactness we abbreviate $\tilde{\omega}' \equiv \omega + \tilde{\omega} - \omega'$. These somewhat formidable expressions can be evaluated in cases of interest, leading to rather intuitive final results, described in the following sections.

### 6.5. Four-frequency correlation functions and quantum advantage

If the exciting pulse is a pure coherent state, such that $\hat{a}(\omega)|\alpha\rangle = \alpha_0 \phi(\omega)|\alpha\rangle$, then the frequency-domain correlation function is:

$$\begin{aligned}
C_{\omega,coh}^{(4)} &= \langle \alpha | \hat{a}^\dagger(\omega')\hat{a}^\dagger(\tilde{\omega}')\hat{a}(\omega)\hat{a}(\tilde{\omega}) | \alpha \rangle \\
&= N^2 \phi^*(\omega')\phi^*(\tilde{\omega}')\phi(\omega)\phi(\tilde{\omega}) \ ,
\end{aligned} \quad (90)$$

where we used that the mean number of photons in the pulse is $N = |\alpha_0|^2$. The temporal amplitude, from Eq.(17), is:

$$A(t) = \alpha_0 \int d\omega \phi(\omega) e^{-i(\omega - \omega_0)t} \ , \quad (91)$$

which acts essentially like a 'classical' optical pulse envelope.

Turning to 'quantum' light, if the exciting pulse is a one-photon pulse, then the correlation function equals zero—there can be no TPA. If the exciting pulse is a two-photon pulse, then the correlation function equals:

$$\begin{aligned}
C_{\omega,\Psi}^{(4)} &= \left\langle \Psi^{(2)} \left| \hat{a}^\dagger(\omega')\hat{a}^\dagger(\tilde{\omega}')\hat{a}(\omega)\hat{a}(\tilde{\omega}) \right| \Psi^{(2)} \right\rangle \\
&= \left\langle \Psi^{(2)} \left| \hat{a}^\dagger(\omega')\hat{a}^\dagger(\tilde{\omega}') \right| vac \right\rangle \left\langle vac \left| \hat{a}(\omega)\hat{a}(\tilde{\omega}) \right| \Psi^{(2)} \right\rangle .
\end{aligned} \quad (92)$$

which is the frequency-domain counterpart of Eq.(78). As we did in Eq.(80), we insert the two-photon component of the state in Eq.(28) into the right-most factor in expression Eq.(92):

$$\langle vac | \hat{a}(\omega)\hat{a}(\tilde{\omega}) | \Psi^{(2)} \rangle = \varepsilon \int d\omega' \int d\tilde{\omega}' \psi(\omega',\tilde{\omega}') \langle vac | \hat{a}(\omega)\hat{a}(\tilde{\omega})\hat{a}^{\dagger}(\omega')\hat{a}^{\dagger}(\tilde{\omega}') | vac \rangle$$
$$= \varepsilon \psi(\omega,\tilde{\omega}) + \varepsilon \psi(\tilde{\omega},\omega) \qquad (93)$$
$$= 2\varepsilon \Psi(\omega,\tilde{\omega}).$$

where we used the delta-function form of the vacuum correlation function $C^{(4)}_{\omega,vac}$ from Eq.(82) and the symmetrized state $\Psi(\omega,\tilde{\omega})$ in Eq.(84). Then, the four-frequency correlation function for the two-photon state is:

$$C^{(4)}_{\omega,\Psi} = 4\varepsilon^2 \Psi^*(\omega',\tilde{\omega}')\Psi(\omega,\tilde{\omega}). \qquad (94)$$

Here we see two essential differences between the classical-state and quantum-state cases:

1) The correlation function for the coherent state is the product of a function of a single variable evaluated at four frequencies, which for the two-photon state it is the product of a function of two variables evaluated at two pairs of frequencies. The two descriptions can be compared by identifying:

$$\Psi(\omega,\tilde{\omega}) \triangleq \phi(\omega)\phi(\tilde{\omega}) \ . \qquad (95)$$

When a function of two variables equals the product of two single-variable functions, we say it is *separable*. Thus, in this case we can identify a coherent state as being identical to a two-photon state in its spectral properties. On the other hand, the quantum state $\Psi(\omega,\tilde{\omega})$ has the possibility to represent correlations of the frequencies of the two photons. These are the entangled photon pairs (EPP), and we will see such a correlation can offer a kind of *quantum advantage* in that it can enhance the probability of TPA.

2) The mean number of photons for the coherent state is $N$, while for the two-photon state the mean number of photons is twice the mean number of pairs, or $2\varepsilon^2$. Thus, from Eqs.(90) and (94), we see that the correlation function and thus the $f$-state probability scales *quadratically* with the mean photon number for the coherent state, while for the two-photon state it scales *linearly* with the mean photon number. This quadratic scaling may offer a second kind of quantum advantage when using a two-photon state to drive TPA in cases where the photon flux is extremely low.

### 6.6. Far-off-resonance approximation

If the exciting field is far from resonance with any intermediate states, as in **Fig. 2a**, we approximate Eq.(89) as:

$$R_{e,e'}^{DQC} \simeq \frac{L_0^{\ 4}}{(-\omega_{fe'} + \omega_0)(\omega_{eg} - \omega_0)} \int d\omega' \int d\omega \int d\tilde{\omega} \frac{\left\langle \hat{a}^\dagger(\omega') \hat{a}^\dagger(\tilde{\omega}') \hat{a}(\omega) \hat{a}(\tilde{\omega}) \right\rangle}{\gamma_{fg} - i\omega_{fg} + i\omega + i\tilde{\omega}}$$

$$R_{e,e'}^{NRP} \simeq \frac{L_0^{\ 4}}{(-\omega_{fe'} + \omega_0)(\omega_{eg} - \omega_0)} \int d\omega' \int d\omega \int d\tilde{\omega} \frac{\left\langle \hat{a}^\dagger(\omega') \hat{a}^\dagger(\tilde{\omega}') \hat{a}(\omega) \hat{a}(\tilde{\omega}) \right\rangle}{\gamma_{ee'} - i\omega_{ee'} + i\omega' - i\omega} \qquad (96)$$

$$R_{e,e'}^{RP} \simeq \frac{-L_0^{\ 4}}{(-\omega_{fe} + \omega_0)(\omega_{eg} - \omega_0)} \int d\omega' \int d\omega \int d\tilde{\omega} \frac{\left\langle \hat{a}^\dagger(\omega') \hat{a}^\dagger(\tilde{\omega}') \hat{a}(\omega) \hat{a}(\tilde{\omega}) \right\rangle}{\gamma_{ee'} - i\omega_{ee'} + i\omega' - i\omega}.$$

where $\omega_0$ is the central frequency of the driving-field spectrum, and we still abbreviate $\tilde{\omega}' \equiv \omega + \tilde{\omega} - \omega'$. We can gain some insight by examining the denominators inside the integrals. For DQC, the denominator is minimized when $\omega + \tilde{\omega} = \omega_{fg}$, that is the two photon's frequencies sum to the two-photon resonance frequency. For NRP and RP, the denominators are minimized when $\omega' - \omega = \omega_{ee'}$, that is the difference of the two photon's frequencies equals either zero (for $e' = e$, meaning the pathway goes through a 'real' population of an intermediate state), or $\omega_{e'e} \neq 0$ (meaning the pathway goes through a 'coherence' between two states $e$ and $e'$).

In this tutorial we focus on the case that the DQC term is two-photon resonant or near-resonant, which means that the NRP and RP terms are far-off resonance in most practical situations. An exception is a molecule such as a dimer consisting of two like monomers in which the two-photon resonance exciting the doubly excited state has frequency very near the monomer single-photon absorption transitions. In such cases, one should consider the limitations imposed by the rotating-wave approximation and consider additional terms in the expansion of the nested commutators in Eq.(68).

In all the following we will assume the condition of far-off-resonance intermediated states. We will show that under this condition the DQC pathway dominates, and that it may be strongly enhanced by time-frequency entanglement.

## 7. Two-photon excitation by coherent states

Here we discuss two-photon excitation by coherent states in several scenarios. The examples serve as baselines for the comparisons to ETPA in later sections.

For excitation by a coherent state, we insert Eq.(90) into the far-off-resonance expressions Eq.(89) to give:

$$R_{e,e'}^{DQC} \simeq \frac{L_0{}^4}{(-\omega_{fe'} + \omega_0)(\omega_{eg} - \omega_0)} \int d\omega' \int d\omega \int d\tilde{\omega} \, \frac{\phi^*(\omega')\phi^*(\tilde{\omega}')\phi(\omega)\phi(\tilde{\omega})}{\gamma_{fg} - i\omega_{fg} + i\omega + i\tilde{\omega}}$$

$$R_{e,e'}^{NRP} \simeq \frac{L_0{}^4}{(-\omega_{fe'} + \omega_0)(\omega_{eg} - \omega_0)} \int d\omega' \int d\omega \int d\tilde{\omega} \, \frac{\phi^*(\omega')\phi^*(\tilde{\omega}')\phi(\omega)\phi(\tilde{\omega})}{\gamma_{ee'} - i\omega_{ee'} + i\omega' - i\omega} \qquad (97)$$

$$R_{e,e'}^{RP} \simeq \frac{-L_0{}^4}{(-\omega_{fe} + \omega_0)(\omega_{eg} - \omega_0)} \int d\omega' \int d\omega \int d\tilde{\omega} \, \frac{\phi^*(\omega')\phi^*(\tilde{\omega}')\phi(\omega)\phi(\tilde{\omega})}{\gamma_{ee'} - i\omega_{ee'} + i\omega' - i\omega} \, ,$$

where, again, $\tilde{\omega}' \equiv \omega + \tilde{\omega} - \omega'$.

## 7.1 Coherent-state DQC pathway

Consider a coherent-state pulse with arbitrary shape and corresponding spectral amplitude $\phi(\omega)$. In the first line of Eq.(97) we change variables to $x = \omega + \tilde{\omega} - 2\omega_0$, $z = \omega - \omega_0$ (and $z' = \omega' - \omega_0$) and obtain:

$$R_{e,e'}^{DQC} \simeq \frac{L_0{}^4}{(-\omega_{fe'} + \omega_0)(\omega_{eg} - \omega_0)} \int dx \, \frac{|K_{coh}(x)|^2}{\gamma_{fg} - i\omega_{fg} + i(2\omega_0 + x)} \, , \qquad (98)$$

where we define:

$$K_{coh}(x) = \int dz \, \phi(\omega_0 + z)\phi(\omega_0 + x - z) \qquad (99)$$

Combining Eqs.(87), (97), and (99), we find for the $f$-state population via the DQC pathway:

$$P_f^{DQC} = N^2 L_0{}^4 \, \Sigma^{(2)} \int dx \, \frac{|K_{coh}(x)|^2}{\gamma_{fg} - i\omega_{fg} + i(2\omega_0 + x)} \; + \; cc \, , \qquad (100)$$

where for convenience we denote:

$$\Sigma^{(2)} = \sum_{e,e'} \frac{\mu_{fe'}\mu_{e'g}\mu_{ge}\mu_{ef}}{(-\omega_{fe'} + \omega_0)(\omega_{eg} - \omega_0)} \, . \qquad (101)$$

$K_{coh}(x)$ is a projection along anti-diagonal lines, as illustrated in **Fig. 3**. It represents the different combinations of frequencies that effectively create excitation near $\omega + \tilde{\omega} = 2\omega_0$.

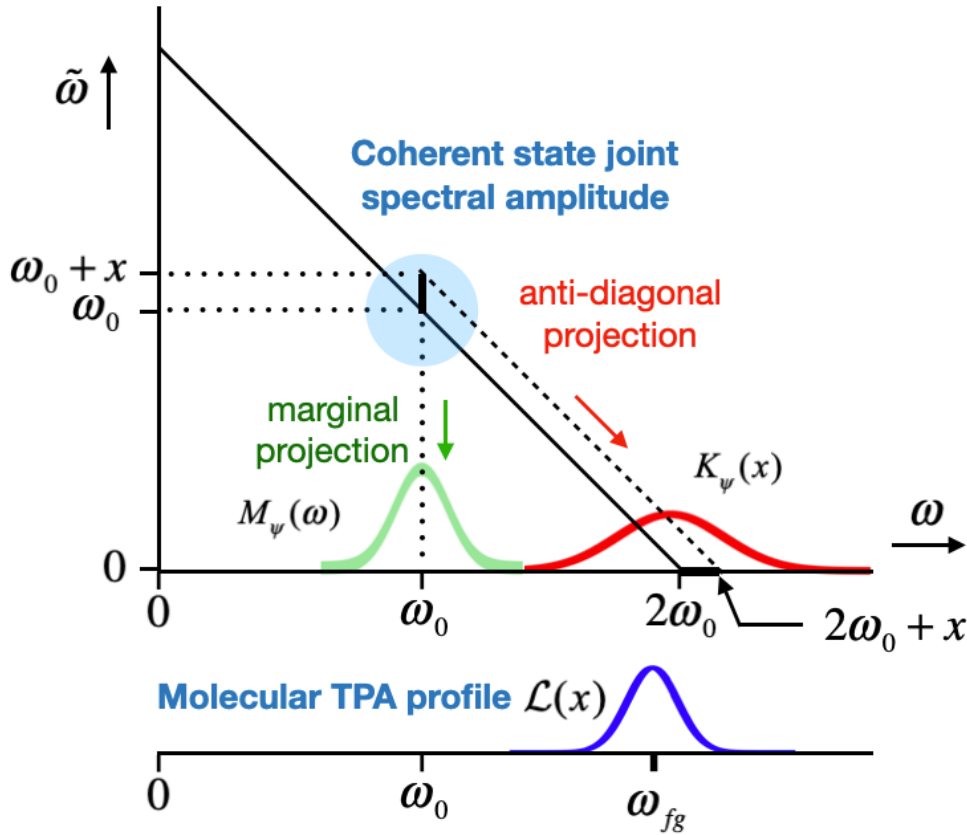

Fig. 7 The anti-diagonal and the marginal (vertical) projections of the JSA, $\psi(\omega, \tilde{\omega})$, for a coherent state. The anti-diagonal projection of the JSA onto the spectral region containing the two-photon absorption profile determines the probability of TPA.

As shown in **Fig. 7**, we also define a *marginal spectrum* for the coherent state as the *vertical* projection onto the $\omega$ axis:

$$M_{coh}(\omega) = \int d\tilde{\omega} \left| \phi(\omega)\phi(\tilde{\omega}) \right|^2$$
$$= \left| \phi(\omega) \right|^2 . \qquad (102)$$

which is identical to the standard spectrum as measured by a grating spectrometer with linear-response detector. In contrast, the anti-diagonally projected spectrum $K_{coh}(x)$ is that 'felt' by the molecule, which acts as a spectrally selective two-photon detector.

### 7.2 Coherent-state exponential pulse: DQC pathway

A nice example that permits an analytical result is TPA excited by a coherent-state pulse in the form of a single-sided exponential, as in Eq.(27)

$$A(t) = \begin{cases} 0, \ t < 0 \\ A(0)\sqrt{2\Gamma}\exp[-(i\omega_0 + \Gamma)t], \ t > 0 \end{cases} . \tag{103}$$

for which the mean total number of photons is $N = |\alpha_0|^2 = |A(0)|^2$ and the spectral amplitude is:

$$\phi(\omega) = \frac{\sqrt{2\Gamma}}{\Gamma - i(\omega - \omega_0)} , \tag{104}$$

with full linewidth equal to $2\Gamma$. The integrals in Eqs.(98), (99) can be carried out to give:

$$K_{coh}(x) = \frac{2\Gamma}{2\Gamma - ix} , \tag{105}$$

and:

$$R_{e,e'}^{DQC} \simeq \frac{L_0^{\ 4}|A(0)|^4}{(-\omega_{fe'} + \omega_0)(\omega_{eg} - \omega_0)} \cdot \frac{\Gamma}{\gamma_{fg} + 2\Gamma + i(2\omega_0 - \omega_{fg})} . \tag{106}$$

We see that in this model the coherent-state bandwidth simply adds to the dephasing line width of the TPA transition. The population created by the DQC pathway, on two-photon resonance, from Eqs.(87) and (106), is:

$$P_f^{DQC} = N^2 L_0^{\ 4} \sum_{e,e'} \frac{M_{fe'eg}}{(-\omega_{fe'} + \omega_0)(\omega_{eg} - \omega_0)} \times \frac{2\Gamma}{\gamma_{fg} + 2\Gamma} + cc . \tag{107}$$

### 7.3 Rectangular coherent state pulse and TPA cross section

An absorption cross section is an effective area that describes the probability per second that a photon (or photons) will be absorbed from a beam with constant flux $F_{coh}$ (photons/s) and given area $A_0$ (m$^2$). To determine the TPA cross section, consider a 'rectangular' coherent-state pulse that suddenly turns on with constant amplitude for a duration $T$, and is zero afterward, as in Eq. (19). The square-normalized spectral amplitude is:

$$\phi(\omega) = \frac{\sin[(\omega - \omega_0)T/2]}{\sqrt{T}(\omega - \omega_0)/2} . \tag{108}$$

The anti-diagonal-projected spectrum, from Eq.(99), is then:

$$K_{coh}(x) = \frac{\sin[xT/2]}{xT/2} . \tag{109}$$

This leads to the result, if the field is two-photon resonant, $\omega_{fg} = 2\omega_0$:

$$P_f^{DQC} = N^2 L_0^{\;4} \Sigma^{(2)} \frac{1}{\gamma_{fg} T}\left(1 + \frac{e^{-\gamma_{fg}T} - 1}{\gamma_{fg} T}\right) \; + \; cc \; . \tag{110}$$

The exponential term here is a turn-on transient. If the exciting field has long duration and thus is quasi-monochromatic, that is spectrally narrow compared to the TPA transition linewidth ($1/T << \gamma_{fg}$), then Eq.(110) becomes:

$$P_f^{DQC} \simeq \frac{N^2}{T}\frac{L_0^{\;4}}{\gamma_{fg}}\Sigma^{(2)} \; + \; cc \; . \tag{111}$$

To write this probability in terms of a cross section, note that the instantaneous, constant photon flux for a 'rectangular' pulse is $F_{coh} = N/T$, therefore $N^2/T = (N^2/T^2)T = F_{coh}^{\;2}T$. We can thus define a two-photon cross section by writing the probability increase per second in terms of a flux density $F_{coh}/A_0$ (photons per $s$ per $m^2$) as:

$$P_f^{DQC}/T = \left(\frac{F_{coh}}{A_0}\right)^2 \sigma^{(2)}, \tag{112}$$

where

$$\begin{aligned}
\sigma^{(2)} &= A_0^{\;2} L_0^{\;4}\frac{1}{\gamma_{fg}}\Sigma^{(2)} \; + \; cc \\
&= \left(\frac{\omega_0}{\hbar\varepsilon_0 nc}\right)^2 \frac{1}{2\gamma_{fg}}\sum_{e,e'}\frac{d_{fe}d_{e'g}d_{ge}d_{ef}}{(-\omega_{fe'}+\omega_0)(\omega_{eg}-\omega_0)},
\end{aligned} \tag{113}$$

where we used $L_0 = (\hbar\omega_0/2\varepsilon_0 ncA_0)^{1/2}$ and $\mu_{ij} = d_{ij}/\hbar = \mathbf{d}_{ij}\cdot\mathbf{e}/\hbar$. This result is precisely the far-off-resonance two-photon cross section derived using second-order time-dependent perturbation theory and averaging over the density of states, as can be seen by comparing the double sum to:

$$\left|\sum_e \frac{d_{ef}d_{ge}}{\omega_{eg}-\omega_0}\right|^2 = \sum_{e'}\frac{d_{fe'}d_{e'g}}{\omega_{e'g}-\omega_0}\sum_e \frac{d_{ef}d_{ge}}{\omega_{eg}-\omega_0} \; , \tag{114}$$

and noting that at TPA resonance $\omega_{e'g}-\omega_0 = -\omega_{fe'}+\omega_0$. Note that to achieve agreement between our result and the conventional one, we had to consider that the pulse is long enough ($T >> 1/\gamma_{fg}$) to allow neglecting the turn-on transient, as is usual in using perturbation theory for determining rates. But we cannot take the pulse too long, because according to this perturbative theory the population increases linearly in time; that is, there is no steady-state limit in this treatment.

For molecules in solution $\sigma^{(2)}$ is typically of the order of 1 to 1,000 GM (where $1\,GM = 10^{-58}\,m^4 s$). [29] Then, for a steady 1-watt laser beam with wavelength 800 nm collimated to an area $5\,\mu m^2$, the flux is $F_{coh} = 4.0\times10^{18}\,photons/s$ and the flux density is $F_{coh}/A_0 = 8\times10^{29}\,photons/m^2 s$, and the expected TPA rate per molecule is about 64/s. For a 1-ns pulse, roughly the upper limit for the applicability of the present perturbation theory, the probability to excite a given molecule is thus $6.4 \times 10^{-8}$. Given that $4.0 \times 10^{9}$ photons pass through the beam area at the molecule's vicinity, we infer an extremely small TPA efficiency per photon per molecule. This low efficiency is one of the motivations for exploring whether time-frequency entanglement of photons can greatly increase this efficiency.

For completeness, Eq.(112) can be generalized for a long arbitrarily shaped, quasi-monochromatic light pulse with a time-dependent flux $F_{coh}(t)$, and a nonzero detuning between $2\omega_0$ and $\omega_{fg}$ to:

$$P_f^{DQC} = \sigma^{(2)} \frac{\gamma_{fg}^2}{\gamma_{fg}^2 + (\omega_{fg} - 2\omega_0)^2} \frac{1}{A_0^2} \int |A(t)|^4 \, dt$$

$$= \sigma^{(2)} \frac{\gamma_{fg}^2}{\gamma_{fg}^2 + (\omega_{fg} - 2\omega_0)^2} \int F_{coh}^2(t) \, dt \; ,$$

(115)

where we used the Fourier relation:

$$\int |A(t)|^4 \, dt = \int d\omega \int d\tilde{\omega} \int d\omega' \, \alpha^*(\omega') \alpha^*(\omega + \tilde{\omega} - \omega') \alpha(\omega) \alpha(\tilde{\omega}) \; .$$

(116)

### 7.4 Gaussian coherent-state pulse: DQC pathway

As another example of TPA, consider a Gaussian-pulse coherent state having a spectrum with bandwidth $\sim \sigma$, i.e., $\alpha(\omega) = \alpha_0 (\sigma^2 / 2\pi)^{-1/4} \exp(-(\omega - \omega_0)^2 / 4\sigma^2)$, and duration $\sim 1/\sigma$. The total number of photons is $N = \int |\alpha(\omega)|^2 \, d\omega = |\alpha_0|^2$. Then we find for the anti-diagonally projected spectrum $K_{coh}(x) = \exp(-x^2 / 8\sigma^2)$ and, for two-photon resonance, $2\omega_0 = \omega_{fg}$, the excitation probability is, from Eq.(100):

$$P_f^{DQC} = N^2 \Sigma^{(2)} L_0^4 \xi(\gamma_{fg} / 2\sigma) \; ,$$

(117)

where $\Sigma^{(2)}$ is defined in Eq.(101) and where $\xi(z) = \exp(+z^2) erfc(z)$, (sometimes denoted as $erfcx(z)$) which has maximum value 1 at $z = 0$ (an ultrashort pulse). For a long, quasi-monochromatic pulse (large $z$), $\xi(z)$ decays to zero as $\sim 1/(\pi^{1/2} z)$. We summarize these two limits:

$$P_f^{coh} = \begin{cases} N^2 \Sigma^{(2)} L_0^4 \dfrac{1}{\sqrt{\pi}} \dfrac{2\sigma}{\gamma_{fg}} & (\gamma_{fg} >> \sigma) \\[2ex] N^2 \Sigma^{(2)} L_0^4 & (\gamma_{fg} << \sigma) \end{cases} \; .$$

(118)

We see, again, the expected quadratic dependence on the number of photons. Note that in the impulsive limit $\sigma \to \infty$ the probability does not depend on $\gamma_{fg}$ or $\sigma$ because for the coherent ('instantaneous-response') nonlinear TPA process the effect of spreading the spectrum over a broader range is compensated by the increasing peak intensity in the time domain, as verified in **Appendix C** also for an arbitrary pulse shape. In the opposite limit $\sigma \to 0$, where the pulse duration tends to infinity, the probability goes to zero because the fixed number of photons are spread over a longer and longer time interval, decreasing the chances for accidental coincidences.

### 7.5 Coherent-state NRP and RP pathways

The two pathways labeled NRP and RP in **Fig. 2** may also contribute to the probability $P_f = Tr\,\hat{\rho}^{(4)}|f\rangle\langle f|$ for exciting the $f$ state. In these pathways, TPA proceeds through a 'step-wise' process through the population of state $e$, $\rho_{gg} \to \rho_{eg} \to \rho_{ee} \to \rho_{fe} \to \rho_{ff}$, [54] or through a coherent process $\rho_{gg} \to \rho_{eg} \to \rho_{ee'} \to \rho_{fe'} \to \rho_{ff}$, in which two intermediate states $e$ and $e'$ are excited coherently.

We focus on the case that the center frequency of the exciting light is near two-photon resonance with the $f$-$g$ transition, that is, $2\omega_0 \simeq \omega_{fg}$, where the rotating-wave approximation is most reliable. We address whether or not the NRP and RP pathways contribute significantly to the $f$-state probability in comparison to the DQC pathway in the case of coherent-state excitation.

Recall that the DQC pathway creates a TPA cross section in exact agreement with the Göppert-Mayer perturbation theory after any initial transients have damped out, as shown in **Sec. 6**. We would like to learn whether the NRP and RP terms together can nevertheless contribute to the cross section under certain conditions.

To evaluate the NRP and RP terms, we write the second and third lines of Eq.(96) as:

$$R_{e,e'}^{NRP} \simeq \frac{L_0^{\;4}}{(-\omega_{fe'}+\omega_0)(\omega_{eg}-\omega_0)}Q_{e,e'}$$
$$R_{e,e'}^{RP} \simeq \frac{-L_0^{\;4}}{(-\omega_{fe}+\omega_0)(\omega_{eg}-\omega_0)}Q_{e,e'} \;\;,$$
$$\tag{119}$$

where, for arbitrary states of light:

$$Q_{e,e'} = \int d\omega' \int d\omega \int d\tilde{\omega}\, \frac{\left\langle \hat{a}^{\dagger}(\omega')\hat{a}^{\dagger}(\omega+\tilde{\omega}-\omega')\hat{a}(\omega)\hat{a}(\tilde{\omega})\right\rangle}{\gamma_{ee'}-i\omega_{ee'}+i\omega'-i\omega} \tag{120}$$

First, we observe from Eq.(119) that for the terms with $e' = e$, so that $\omega_{fe'} = \omega_{fe}$, we have $R_{e,e}^{NRP} + R_{e,e}^{RP} = 0$ in this far-off-resonance situation where dephasing is unimportant. Thus, if there is only a single relevant intermediate state, then the NRP and RP pathways cancel and do not contribute. If the dephasing line width $\gamma_{fe}$ becomes significant, as in the original Eq.(89), it can 'destroy' the near-perfect cancellation of $R_{e,e'}^{NRP}$ and $R_{e,e'}^{RP}$, leading to active pathways for TPA that do not exist in the absence of fast dephasing. Study of these pathways requires a more detailed analysis than presented on this tutorial. This "destruction of destructive interference" is similar to that studied in different contexts in, for example, [50, 51, 52] and reviewed in [53, 54].

So, in the general case only pathways with $e' \neq e$, in which two intermediate states $e$ and $e'$ are excited coherently, contribute to the $f$-state probability. To analyze cases with $e' \neq e$, in Eq.(120) we change variables to $y = \omega - \omega'$ and write it in the form:

$$Q_{e,e'} = \int dy \frac{G(y)}{\gamma_{ee'} - i(\omega_{ee'} + y)}, \tag{121}$$

where, for a coherent state:

$$G(y) = \int d\omega \int d\tilde{\omega} \left\langle \hat{a}^\dagger(\omega - y)\hat{a}^\dagger(\tilde{\omega} + y)\hat{a}(\omega)\hat{a}(\tilde{\omega}) \right\rangle. \tag{122}$$

This expression is valid for excitation by coherent states or two-photon states. For a coherent state, it becomes:

$$G_{coh}(y) = |\alpha_0|^4 \int d\omega \int d\tilde{\omega} \phi^*(\omega - y)\phi^*(\tilde{\omega} + y)\phi(\omega)\phi(\tilde{\omega}). \tag{123}$$

If we define an effective 'two-photon amplitude' for the coherent state as $\Psi_{coh}(\omega, \tilde{\omega}) = \phi(\omega)\phi^*(\tilde{\omega})$, then we have:

$$G_{coh}(y) = |\alpha_0|^4 \int d\omega \int d\tilde{\omega} \Psi_{coh}^*(\omega - y, \tilde{\omega} + y)\Psi_{coh}(\omega, \tilde{\omega}). \tag{124}$$

This integral is an autocorrelation of a two-dimensional function $\Psi_{coh}(\omega, \tilde{\omega})$, with the shift being along the anti-diagonal direction in the $\omega, \omega'$ plane. It is not the same as the anti-diagonal projection $K_{coh}(x)$ that appears in the analogous formula, Eq.(99), for the DQC pathway.

### 7.6 Exponential coherent state pulse: NRP and RP pathways

Here we consider under what conditions the NRP and RP contributions to the $f$-state probability cancel exactly or approximately. We use Eqs.(87) and (119) to write their sum:

$$P_f^{NRP} + P_f^{RP} = \sum_{e,e'} M_{fe'eg} \left( R_{e,e'}^{NRP} + R_{e,e'}^{RP} \right) + cc, \tag{125}$$

where:

$$\left( R_{e,e'}^{NRP} + R_{e,e'}^{RP} \right) = L_0^4 \left( \frac{1}{(-\omega_{fe'} + \omega_0)(\omega_{eg} - \omega_0)} - \frac{1}{(-\omega_{fe} + \omega_0)(\omega_{eg} - \omega_0)} \right) Q_{e,e'}$$

$$= \frac{L_0^4}{(\omega_{eg} - \omega_0)} \frac{\omega_{ee'}}{(-\omega_{fe'} + \omega_0)(-\omega_{fe} + \omega_0)} Q_{e,e'}. \tag{126}$$

Thus, terms in which two intermediate states are degenerate ($\omega_{ee'} = 0$) do not contribute.

To compare with the DQC term, again consider excitation by a coherent-state pulse in the form of a single-sided exponential, as in Eq.(103). Then $G_{coh}(y) = 4\Gamma^2 / (y^2 + 4\Gamma^2)$, which gives:

$$Q_{ee'} = |\alpha_0|^4 \frac{\Gamma(\gamma_{ee'} + 2\Gamma + i\omega_{ee'})}{(\gamma_{ee'} + 2\Gamma)^2 + \omega_{ee'}^2} \ , \tag{127}$$

Consider the special case in which the product of the four dipole matrix elements $M_{fe'eg}$ is real, the plausibility of which is discussed in **Appendix B**. Then the sum of NRP and RP contributions to the $f$-state probability depend only on the real part of $R_{e,e'}^{NRP} + R_{e,e'}^{RP}$:

$$\text{Re}\left(R_{e,e'}^{NRP} + R_{e,e'}^{RP}\right) = \frac{L_0^4 A^2(0)}{(-\omega_{fe'} + \omega_0)(\omega_{eg} - \omega_0)} \cdot \frac{\omega_{ee'}}{(-\omega_{fe} + \omega_0)} \cdot \frac{\Gamma(\gamma_{ee'} + 2\Gamma)}{(\gamma_{ee'} + 2\Gamma)^2 + \omega_{ee'}^2} \ . \tag{128}$$

Compare this result to the corresponding result for DQC when on TPA resonance, from Eq.(106):

$$R_{e,e'}^{DQC} = \frac{L_0^4 A^2(0)}{(-\omega_{fe'} + \omega_0)(\omega_{eg} - \omega_0)} \frac{\Gamma}{(\gamma_{fg} + 2\Gamma)} \ . \tag{129}$$

The first factors of these two expressions are identical. The magnitude of the middle term in Eq.(128), $\omega_{ee'} / (-\omega_{fe} + \omega_0)$, is of the order of unity, or much smaller. The third term has two limiting behaviors:

$$\frac{\Gamma(\gamma_{ee'} + 2\Gamma)}{(\gamma_{ee'} + 2\Gamma)^2 + \omega_{ee'}^2} \rightarrow \begin{cases} \dfrac{\Gamma}{\gamma_{ee'} + 2\Gamma} \ , \ (\gamma_{ee'} + 2\Gamma)^2 >> \omega_{ee'}^2 \\[3mm] \dfrac{\Gamma(\gamma_{ee'} + 2\Gamma)}{\omega_{ee'}^2} , \ \omega_{ee'}^2 >> (\gamma_{ee'} + 2\Gamma)^2 \end{cases} \tag{130}$$

Although these expressions are rather complex, we conclude that the NRP+RP term can be comparable to the DQC term under some conditions. On the other hand, it will be much smaller than the DQC term if $\omega_{ee'}$ is much smaller in magnitude than $(-\omega_{fe} + \omega_0)$ and $\gamma_{ee'}$ is comparable to or greater than $\gamma_{fg}$.

Going beyond the special case that $M_{fe'eg}$ is real requires more complicated analysis, which we avoid in this tutorial.

### 7.7 Coherent-state rectangular pulse: NRP and RP pathways

Consider a rectangular coherent-state pulse exciting the NRP and RP pathways. To evaluate Eq.(121) we repeat the method of calculation in **Sec. 7.3**, first using Eq.(108) in Eq.(123), which gives $G_{coh}(y) = \sin^2(yT/2)/(yT/2)^2$. Then we find:

$$Q_{e,e'} = |\alpha_0|^4 \frac{\gamma_{ee'} + i\omega_{ee'}}{(\gamma_{ee'}^2 + \omega_{ee'}^2)T} \left( 1 + \frac{e^{-(\gamma_{ee'} - i\omega_{ee'})T} - 1}{(\gamma_{ee'} - i\omega_{ee'})T} \right). \tag{131}$$

If the rectangular pulse is much longer than the inverse dephasing rate, $T \gg 1/\gamma_{ee'}$, we can neglect the second term, which is an oscillating turn-on transient.

## 8. Two-photon excitation by entangled two-photon states

We now come to a most interesting case, which is the primary motivation for this tutorial—TPA with entangled photon pairs in the isolated-pair regime. As discussed in the Introduction, an important question is how large can be the enhancement of TPA by time-frequency entanglement in the EPP state? We expect such an enhancement if the photon pairs' nondeterministic frequencies are correlated and sum to the TPA resonance frequency, even if each has an average spectrum that is much broader than the TPA transition line width. Again, we focus on the case that the intermediate molecular states are far from resonance. We find that large enhancement by time-frequency entanglement is possible. [20, 21] And we show that the DQC pathway dominates, as we showed earlier for coherent state excitation under certain conditions.

### 8.1 Two-photon-state DQC pathway

The probability to excite the $f$ state via the far-off-resonance DQC pathway is given in general by Eqs.(87), (94), (96), and (101) as:

$$P_f^{DQC} = 4\varepsilon^2 \Sigma^{(2)} L_0^4 \int d\omega' \int d\omega \int d\tilde{\omega} \frac{\Psi^*(\omega', \tilde{\omega}')\Psi(\omega, \tilde{\omega})}{\gamma_{fg} - i\omega_{fg} + i\omega + i\tilde{\omega}} + cc, \tag{132}$$

where, again, $\tilde{\omega}' \equiv \omega + \tilde{\omega} - \omega'$.

Changing variables to $x = (\omega + \tilde{\omega} - 2\omega_0), z = \omega - \omega_0$ (and $z' = \omega' - \omega_0$), analogously to the coherent-state case, we find:

$$P_f^{DQC} = 4\varepsilon^2 L_0^4 \Sigma^{(2)} \int dx \frac{|K_\psi(x)|^2}{\gamma_{fg} - i\omega_{fg} + i(2\omega_0 + x)} + cc, \tag{133}$$

where:

$$K_\psi(x) = \int dz\, \Psi(\omega_0 + z, \omega_0 + x - z). \tag{134}$$

As for the coherent state result in Eq.(99), $K_\psi(x)$ is an anti-diagonal projection of the JSA (or two-photon amplitude) onto the spectral region around the two-photon resonance. $K_\psi(x)$ has the same form as the coherent-state DQC result with two replacements: $\phi(\omega)\phi(\tilde{\omega}) \rightarrow \Psi(\omega,\tilde{\omega})$ and $N^2 \rightarrow 4\varepsilon^2$. That is, the result for EPP resembles the coherent-state result but with a generalized spectral dependence and with linear instead of quadratic photon-flux dependence. While the coherent state takes the form of a separable two-photon state without frequency correlations, the EPP state is a pure state that can be nonseparable, that is $\Psi(\omega,\tilde{\omega}) \neq \phi(\omega)\phi(\tilde{\omega})$, indicating frequency correlations (entanglement). In particular, the JSA may be much narrower in the direction of the diagonal frequency axis than the anti-diagonal axis, as shown in **Fig. 8**. In this case the frequencies of the two photons are anticorrelated, and, because the state is a pure state, entangled. Such correlations can enhance the rate. And, while the coherent state relies on 'accidental' coincidences for photons to arrive together at the molecule, the EPP state has photons always arriving in pairs, giving linear scaling with flux (proportional to $\varepsilon^2$), as discussed in Sec. **6.5**.

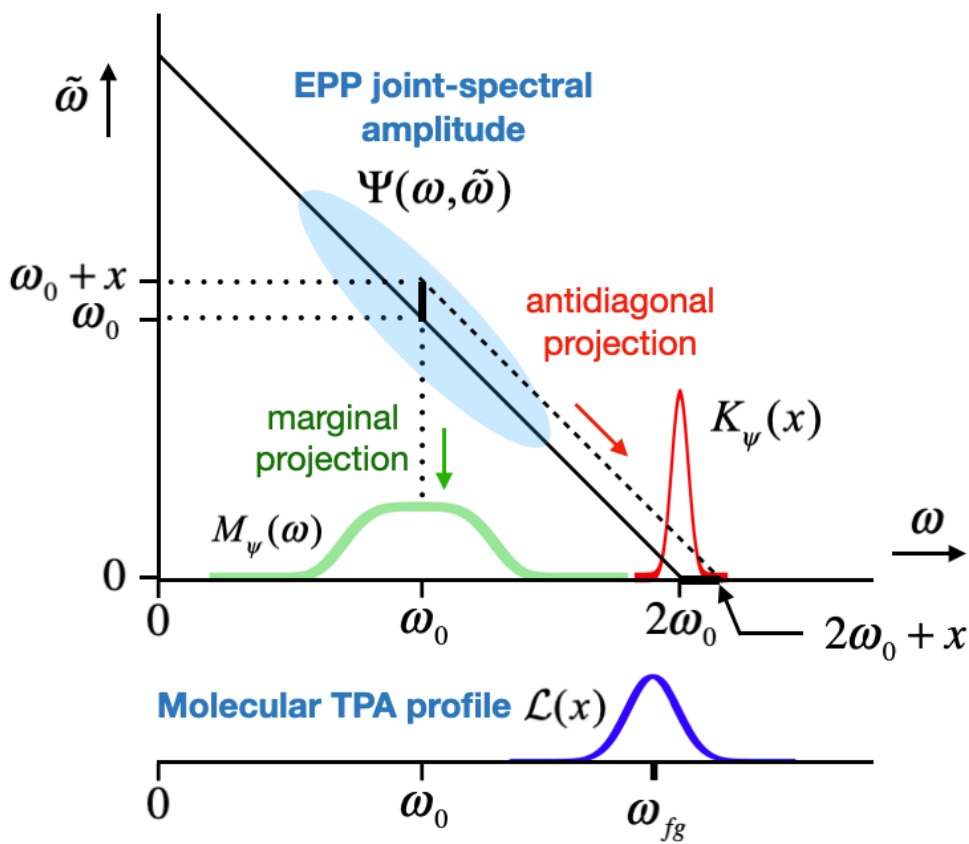

Fig. 8 The marginal (vertical) and anti-diagonal projections of the JSA, $\Psi(\omega,\tilde{\omega})$, for a time-frequency entangled two-photon state. The anti-diagonal projection of the JSA onto the spectral region containing the two-photon absorption profile determines the probability of TPA.

We consider two cases.

### 8.2 Separable two-photon state: DQC pathway

An ultrashort pump pulse together with a particular phase-matching condition of Type-I SPDC can create a separable (factorable) JSA, as in [55]. In this case there is no spectral entanglement. Then because of symmetry, $\psi(\omega, \tilde{\omega}) = \phi_0(\omega)\phi_0(\tilde{\omega})$, which gives:

$$P_f^{DQC} = 4\varepsilon^2 L_0^4 \Sigma^{(2)} \int dx \frac{\left|K_{sep}(x)\right|^2}{\gamma_{fg} - i\omega_{fg} + i(2\omega_0 + x)} + cc \;, \tag{135}$$

where $K_{sep}(x)$ is given by:

$$K_{sep}(x) = \int dz \, \phi_0(\omega_0 + z)\phi_0(\omega_0 + x - z) \;. \tag{136}$$

This expression has precisely the same spectral dependence as for the coherent-state pulse in Eq.(100), and thus in this case, although the state is 'non-classical', there is no enhancement of TPA through spectral correlation. There can still be enhancement by photon number correlation.

### 8.3 Anti-diagonally separable two-photon states: DQC pathway

An important example of time-frequency entanglement is EPP generated by Type-I SPDC using a narrow-band pump pulse with duration $T$ that is long compared to the inverse of the phase-matching bandwidth. Energy conservation localizes the JSA $\Psi(\omega, \tilde{\omega})$ along the antidiagonal, $\tilde{\omega} + \omega = \omega_P$. Properly designed, such a state maximizes the enhancement of TPA by time-frequency EPP. [20, 21]

For degenerate SPDC the JSA has a single peak at the frequency where both $\omega, \tilde{\omega}$ equal $\omega_0 = \omega_P / 2$. In the absence of dispersion, which creates phase correlations, we can model the JSA as the product of narrow and broad functions, $\psi_N(\omega)$ and $\psi_B(\omega)$ respectively, centered at $\omega = \tilde{\omega} = \omega_P / 2$ and oriented along diagonal and antidiagonal axes in the $(\omega, \tilde{\omega})$ plane, respectively. The width of $\psi_N(\omega)$ is the linewidth (inverse duration) of the pump pulse, and the linewidth of $\psi_B(\omega)$ is roughly $\sqrt{2}$ times the spectral width of the EPP, set by the phase-matching conditions. Then:

$$\Psi(\omega, \tilde{\omega}) = \psi_B(\frac{\omega - \tilde{\omega}}{2})\psi_N(\omega + \tilde{\omega} - \omega_P), \tag{137}$$

where, by state symmetry it is required that $\psi_B(-x) = \psi_B(x)$, and both functions are square-normalized in $dx = dx / 2\pi$. Note that while we call such a state 'anti-diagonally separable', it is an entangled state with regard to the frequencies of the two photons.

Then we find:

$$K_\psi(x) = \psi_N(x) \int dz \, \psi_B(z) \, . \tag{138}$$

This gives the probability:

$$P_f^{DQC} = 4\varepsilon^2 L_0^{\ 4} \Sigma^{(2)} \left| \int dz \, \psi_B(z) \right|^2 \int dx \, \frac{\left| \psi_N(x) \right|^2}{\gamma_{fg} - i\omega_{fg} + i(2\omega_0 + x)} \; + \; cc \, . \tag{139}$$

This equation shows that it is the spectrally narrow function $\psi_N(x)$ that effectively drives the molecular transition as a result of the anticorrelation of EPP frequencies.

### 8.4 Gaussian EPP pulse: DQC pathway and EPP enhancement

The above integral can be evaluated by assuming Gaussian forms (valid for Type-0 or Type-I SPDC in the case of a long narrow-band pump pulse):

$$\begin{aligned} \psi_N(x) &= (\sigma_N^{\ 2} / 2\pi)^{-1/4} \exp(-x^2 / 4\sigma_N^{\ 2}) \\ \psi_B(x) &= (\sigma_B^{\ 2} / 2\pi)^{-1/4} \exp(-x^2 / 4\sigma_B^{\ 2}) \end{aligned} , \tag{140}$$

where we impose the condition $\sigma_N < \sigma_B$. The 'narrow' width $\sigma_N$ equals the spectral width of the laser pulse driving the SPDC, while the 'broad' width $\sigma_B$ is determined by phase matching. [55]  In this case:

$$K_\psi(x) = \sqrt{\frac{2\sigma_B}{\sigma_N}} \exp[-x^2 / (2\sigma_N^{\ 2})] \tag{141}$$

which leads to the probability:

$$P_f^{DQC,EPP} = 4\varepsilon^2 L_0^{\ 4} \Sigma^{(2)} \frac{2\sigma_B}{\sigma_N} \xi\left( \gamma_{fg} / \sqrt{2}\sigma_N \right) , \tag{142}$$

where $\xi(z) = \exp(+z^2) erfc(z)$, which we also encountered for the Gaussian shaped coherent-state pulse in **Sec. 7.4**. A related expression was given in [6]. In two limits this becomes:

$$P_f^{EPP} = \begin{cases} 4\varepsilon^2 L_0^{\ 4} \Sigma^{(2)} \dfrac{\sqrt{2}}{\sqrt{\pi}} \dfrac{2\sigma_B}{\gamma_{fg}} & (\gamma_{fg} \gg \sigma_N) \\[3mm] 4\varepsilon^2 L_0^{\ 4} \Sigma^{(2)} \dfrac{2\sigma_B}{\sigma_N} & (\gamma_{fg} \ll \sigma_N) \end{cases} . \tag{143}$$

The factor $\sigma_B / \gamma_{fg}$ in the first of the two expressions in Eq.(143) can be interpreted as the number of temporal modes in the EPP state that impinge on the molecule in the molecular coherence time $1/\gamma_{fg}$, which in this case is much smaller than the EPP pulse duration $(1/\gamma_{fg} << 1/\sigma_N)$. Comparing this result with the comparable one in the first line of Eq.(118), we see that the EPP result differs from the coherent-state result by a factor roughly equal to $(4\varepsilon^2 / N^2)(\sigma_B / \sigma)$. The first of these factors is the ratio of photon fluxes and the second is the ratio of the bandwidths of the excitation pulses. Consider that the photon fluxes are equal, which requires them to be small enough that in the case of EPP there are no overlapping pairs on average. Then we see that the factor $(\sigma_B / \sigma)$ can be much greater than one if the EPP is generated in a broad band containing many time-frequency modes. In contrast, the coherent state pulse is a single time-frequency mode, which contains in this case a single pair of photons on average.

The second of these two expressions, for $\gamma_{fg} << \sigma_N$, is in the impulsive limit with respect to the correlation duration $(\sim 1/\sigma_B)$ of the EPP wave packet. In this Gaussian model of the JSA, the ratio $\sigma_B / \sigma_N$ is a measure of the number of temporal (time-frequency) modes in the EPP state that impinge on the molecule in a single pulse with duration $\sim 1/\sigma_N$, and thus is a measure of entanglement. [56] Given that the EPP always arrive together within a time $1/\sigma_B$, regardless of the duration ($1/\sigma_N$) of the pulse, the probability for TPA is enhanced by this factor relative to a narrow-band coherent pulse of the same duration, wherein the photons arrive independently.

We note that the Gaussian approximation can also be applied to Type-II SPDC, with similar conclusions for the long-pulse case.

### 8.5 Exponential EPP pulse: DQC pathway

For later comparison with the NRP and RP pathways, consider modeling the DQC term with an EPP pulse spectrum given by a complex Lorentzian for the narrow function and a gaussian for the broad function. In the spectral domain we have:

$$\begin{aligned}
\psi_N(x) &= \sqrt{2\Gamma} / (\Gamma - ix) \\
\psi_B(x) &= (\sigma_B^2 / 2\pi)^{-1/4} \exp(-x^2 / 4\sigma_B^2)
\end{aligned} \quad , \tag{144}$$

This model has a single-sided exponential pulse shape in the time domain, as in Eq.(103). Then:

$$\int dx \frac{\left|\psi_N(x)\right|^2}{\gamma_{fg} - i\omega_{fg} + i(2\omega_0 + x)} = \frac{1}{\gamma_{fg} + \Gamma + i(2\omega_0 - \omega_{fg})} \tag{145}$$

and:

$$\left|\int dz\, \psi_B(z)\right|^2 = \sqrt{2/\pi}\, \sigma_B \;. \tag{146}$$

On TPA resonance ($2\omega_0 - \omega_{fg}$) this gives, upon writing $\Sigma^{(2)}$ explicitly:

$$P_f^{DQC} = 4\varepsilon^2 L_0^4 \sum_{e,e'} \frac{M_{fe'eg}}{(-\omega_{fe'} + \omega_0)(\omega_{eg} - \omega_0)} \frac{2\sqrt{2}}{\sqrt{\pi}} \frac{\sigma_B}{\gamma_{fg} + \Gamma}.$$ (147)

This result has limits similar to those in Eq.(143), with $\Gamma$ and $\sigma_B$ playing similar roles. The explanation of the potentially very large enhancement by EPP spectral correlations are similar in this case, showing that enhancement occurs for a variety of pulse shapes.

### 8.6 NRP and RP pathways excited by two-photon states

Here we address whether or not the NRP and RP pathways are significantly enhanced by time-frequency entanglement of the exciting photon pairs. One might expect the answer to be no, because these pathways are impacted by dephasing processes that could disrupt the delicate anticorrelation of the photon pair's frequencies.

In the far-off-resonance case we have for the $f$-state probability, using Eqs.(87) and (96):

$$P_f^{NRP} \simeq L_0^4 \sum_{e,e'} \frac{M_{fe'eg}}{(-\omega_{fe'} + \omega_0)(\omega_{eg} - \omega_0)} Q_{e,e'} + cc$$

$$P_f^{RP} \simeq L_0^4 \sum_{e,e'} \frac{-M_{fe'eg}}{(-\omega_{fe} + \omega_0)(\omega_{eg} - \omega_0)} Q_{e,e'} + cc.$$ (148)

where, as in Eq.(121):

$$Q_{e,e'} = \int d y \frac{G_\Psi(y)}{\gamma_{ee'} - i(\omega_{ee'} + y)},$$ (149)

and, using Eq.(94):

$$G_\Psi(y) = \int d\omega \int d\tilde{\omega} \left\langle \hat{a}^\dagger(\omega - y)\hat{a}^\dagger(\tilde{\omega} + y)\hat{a}(\omega)\hat{a}(\tilde{\omega}) \right\rangle$$

$$= 4\varepsilon^2 \int d\omega \int d\tilde{\omega} \, \Psi^*(\omega - y, \tilde{\omega} + y)\Psi(\omega, \tilde{\omega}).$$ (150)

We wish to compare the NRP and RP contributions Eq.(148) to the DQC contribution Eq.(139) in the case that the JSA equals the product of narrow and broad functions, $\psi_N(\omega)$ and $\psi_B(\omega)$ respectively, centered at $\omega = \tilde{\omega} = \omega_P / 2$ and oriented along diagonal and antidiagonal axes in the $(\omega, \tilde{\omega})$ plane, as in Eq.(137). Then, changing variables to $x = \omega + \tilde{\omega} - 2\omega_0 \equiv \omega_D - 2\omega_0$, $z = (\omega - \omega_0)/2 \equiv \omega_A$, gives for G(y):

$$G_{\Psi}(y) = 4\varepsilon^2 \int d\omega_A \int d\omega_D \left| \psi_N(\omega_D - \omega_P) \right|^2 \psi^*_B(\omega_A - y)\psi_B(\omega_A)$$

$$= 4\varepsilon^2 \int d\omega_A \ \psi^*_B(\omega_A - y)\psi_B(\omega_A) \ . \tag{151}$$

The narrow function has dropped out, meaning the NRP and RP terms are independent of pulse duration in this scenario, which includes a single EPP with entanglement time determined by the broad function. And we see that the autocorrelation of the two-dimensional function $\Psi_{coh}(\omega, \tilde{\omega})$ along the anti-diagonal direction in the $\omega, \omega'$ plane (as in Eq.(124)) has been converted to an autocorrelation of a one-dimensional function $\psi_B(\omega_A)$. Again, it is not the same as the anti-diagonal projection $K_{coh}(x)$ that appears in the analogous formula, Eq.(99), for the DQC pathway.

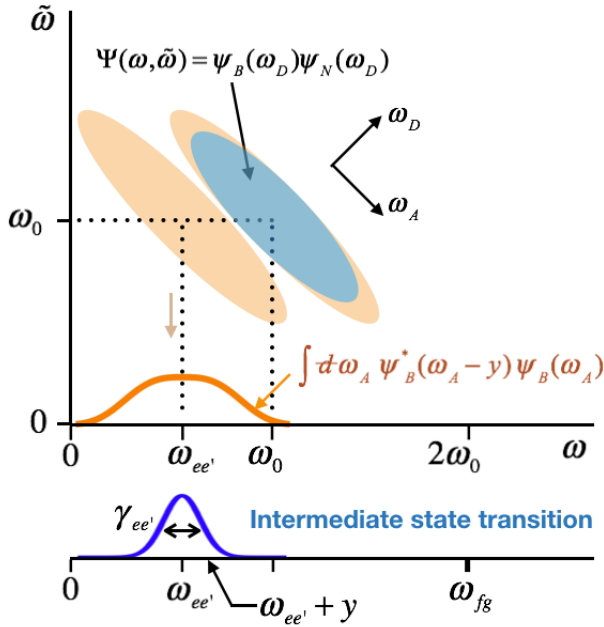

Fig. 9 The contributions of the NRP and RP pathways are determined by an anti-diagonal autocorrelation of the JSA, $\Psi(\omega, \tilde{\omega})$, reduced to a one-dimensional integral along the antidiagonal direction, then multiplied by the intermediate-state $e,e'$ line shape.

The interpretation of Eq.(149) with Eq.(151) and the accompanying illustration in **Fig. 9** is that the population damping rate ($\gamma_{ee}$) of the intermediate state or the dephasing rate ($\gamma_{ee'}$) of the intermediate-state coherence acts as a filter. This filter limits or 'windows' the range of frequencies in the exciting field that contribute to exciting TPA via the NRP and RP pathways through the $\omega_{ee'}$ coherence. In contrast, for the DQC pathway, illustrated in **Fig. 8**, there is no such windowing behavior because the pathway bypasses the intermediate-state populations and coherences.

In contrast to the DQC contribution, which is dominated by the narrow part of the JSA, the NRP and RP contribution are determined by the broad part of the JSA. For this reason, we don't expect enhancement of the NRP and RP processes by spectral correlation.

To find the combined contribution of the NRP and RP probabilities, we sum Eqs.(148) and evaluate the result at the TPA resonance frequency $\omega_0 = \omega_{fg}/2$, to give:

$$P_f^{NRP} + P_f^{RP} = L_0^{\,4} \sum_{e,e'} \frac{M_{fe'eg}}{(\omega_{eg} - \omega_0)(-\omega_{fe'} + \omega_0)} \cdot \frac{\omega_{ee'}}{(-\omega_{fe} + \omega_0)} Q_{e,e'} \ + \ cc \ . \qquad (152)$$

To achieve a direct comparison with the DQC term in a particular scenario, consider again modeling the EPP pulse spectrum by a complex Lorentzian for the narrow function and a gaussian for the broad function, as in Eq.(144). Then the narrow function drops out and the broad function determines $G_\Psi(y)$ in Eq.(151) to be $G_\Psi(y) = 4\varepsilon^2 \exp\left(-y^2/8\sigma_B^{\,2}\right)$, leading to:

$$Q_{e,e'} = 4\varepsilon^2 \int dy \, \frac{\exp\left(-y^2/8\sigma_B^{\,2}\right)}{\gamma_{ee'} - i(\omega_{ee'} + y)} \ . \qquad (153)$$

Rather than attempt this integral in closed form, consider two limits of EPP bandwidth relative to $\gamma_{ee'}$:

$$\begin{aligned}
Q_{e,e'} &\approx \frac{4\varepsilon^2}{\gamma_{ee'} - i\omega_{ee'}} \int dy \exp\left(-y^2/8\sigma_B^{\,2}\right) \\
&= \frac{\sqrt{2}\sigma_B}{\sqrt{\pi}} \frac{4\varepsilon^2}{\gamma_{ee'} - i\omega_{ee'}}, \ \gamma_{ee'} \gg \sigma_B \ , \\
Q_{e,e'} &= \int dy \, \frac{4\varepsilon^2}{\gamma_{ee'} - i(\omega_{ee'} + y)} = \frac{4\varepsilon^2}{2}, \ \sigma_B \gg \gamma_{ee'} \ .
\end{aligned} \qquad (154)$$

We write the sum of probabilities:

$$P_f^{NRP} + P_f^{RP} =$$

$$4\varepsilon^2 L_0^{\,4} \sum_{e,e'} \frac{M_{fe'eg}}{(\omega_{eg} - \omega_0)(-\omega_{fe'} + \omega_0)} \cdot \frac{\omega_{ee'}}{(-\omega_{fe} + \omega_0)} \left\{ \begin{array}{ll} \dfrac{1}{\gamma_{ee'} - i\omega_{ee'}} \dfrac{\sqrt{2}\sigma_B}{\sqrt{\pi}}, & \gamma_{ee'} \gg \sigma_B \\[2mm] \dfrac{1}{2}, & \sigma_B \gg \gamma_{ee'} \end{array} \right\} \ + \ cc \qquad (155)$$

If $M_{fe'eg}$ is real, then:

$$P_f^{NRP} + P_f^{RP} =$$

$$4\varepsilon^2 L_0^{\,4} \sum_{e,e'} \frac{M_{fe'eg}}{(\omega_{eg} - \omega_0)(-\omega_{fe'} + \omega_0)} \cdot \frac{\omega_{ee'}}{(-\omega_{fe} + \omega_0)} \left\{ \begin{array}{ll} \dfrac{\gamma_{ee'}^{\,2}}{\gamma_{ee'}^{\,2} + \omega_{ee'}^{\,2}} \dfrac{2\sqrt{2}}{\sqrt{\pi}} \dfrac{\sigma_B}{\gamma_{ee'}} \ll 1, & \sigma_B \ll \gamma_{ee'} \\[2mm] 1 \, , & \sigma_B \gg \gamma_{ee'} \end{array} \right\}$$

$$(156)$$

Observing that $\gamma_{ee'}^{\;2}/(\gamma_{ee'}^{\;2}+\omega_{ee'}^{\;2})\le 1$, and that the factor $\omega_{ee'}/(-\omega_{fe}+\omega_0)$ has a magnitude is of order unity or less, we can compare this result in either limit with Eq.(147) for that of the DQC pathway under the same form of EPP excitation field. We see that the DQC-pathway probability is greater than that for the NRP+RP pathways by at least a factor roughly equal to:

$$\frac{2\sigma_B}{\gamma_{fg}+\Gamma},\tag{157}$$

which can be much greater than unity if the number of temporal modes, $\sigma_B/\Gamma$, in the EPP pulse is large.

To emphasize this most important conclusion, we have shown that whereas the DQC pathway can be greatly enhanced relative to the coherent-state case, as in Eqs.(143) and (147), the NRP+RP pathways collectively are not enhanced by the spectral anticorrelations (time-frequency entanglement) of the EPP field. Thus, for EPP excitation, the DQC is predicted to be dominant.

### 8.7 Effect of dispersion on TPA by EPP

The frequency-dependent refractive index of optical elements that light passes through before reaching the sample is known to broaden ultrashort pulses temporally and reduce their effectiveness in nonlinear optical processes. To account for such effects in EPP-driven TPA, we incorporate dispersive propagation into the two-photon JSA by replacing: [20]

$$\Psi(\omega,\tilde{\omega})\to\Psi(\omega,\tilde{\omega})\exp[i(D/2)(\omega^2+\tilde{\omega}^2)]\;,\tag{158}$$

where $D$ is the second-order (group delay) dispersion of the transmitting optical system.

As an example, we insert this expression into Eq.(132) for the DQC term excited by EPP. Then using the Gaussian forms in Eq.(140) (valid for long SPDC pump) leads to:

$$K_\psi(x)=\frac{2\sigma_B\exp[-x^2/(2\sigma_N^{\;2})]}{\sigma_N\sqrt{1+16\,D^2\sigma_B^{\;4}}}\;.\tag{159}$$

Comparing to Eq.(141), we see the sole effect of second-order dispersion is to replace $\sigma_B$ by:

$$\tilde{\sigma}_B=\frac{\sigma_B}{\sqrt{1+16\,D^2\sigma_B^{\;4}}}\;.\tag{160}$$

The $x$ dependence of $K_\psi(x)$ is not altered; there is only an overall decrease of magnitude of $K_\psi(x)$ and thus a decrease of the TPA probability.

## 9. Experimental challenges

Regarding the implementation of experiments on ETPA, it is important to consider three issues: 1) What are the experimental signatures that can provide indisputable evidence for ETPA? 2) What are potential reasons one might miss observing ETPA accidentally? 3) What are possible reasons classical signals might be misidentified as evidence for ETPA? Here we follow in part the outline given in the Supplemental Information part of [17], which is based in part on [18, 20].

Experimental signatures that can provide indisputable evidence for ETPA include a combination of a) linear scaling with optical flux incident on the sample (but this alone is not sufficient as several classical processes can mimic this), b) quadratic dependence on linear loss between the SPDC source and the sample, and c) experimental verification that the flux being measured at the sample consists of photon pairs (by coincidence counting of photons in the sample volume when the sample is removed).

Potential reasons one might miss observing ETPA accidentally include a) insufficient spatial overlap of photon pairs in the sample, b) linear dispersion that broadens the EPP correlation time and reduces the effectiveness of ETPA, c) detector saturation and dead-time effects, d) insufficient EPP flux and/or fluorescence collection efficiency, e) reabsorption of fluorescence in the molecular sample.

Possible reasons classical signals might be misidentified as evidence for ETPA include: a) observation of linear scaling of signal with optical flux incident on the sample without performing the other checks mentioned above, b) presence of low-lying resonant intermediate states not recognized for the sample being used, for example such states created by molecular aggregation.

## 10. Summary and Discussion

Given the challenges cited in the previous section, the scientific community that is working to develop ETPA as a tool for quantum-enhanced spectroscopy and imaging of molecular and biological samples is still struggling to identify the techniques and conditions under which such a 'quantum advantage' can be achieved. While several experiments have presented evidence that ETPA in molecules does provide such advantages, other experiments, as well as the theory summarized here, have called those conclusions into question.

We have reviewed the quantum optics theory needed for incorporating entangled quantum states of light into the theory of two-photon absorption by atoms or molecules. The density matrix (or operator) in fourth-order perturbation theory plays a central role because it can describe damping and dephasing of the states and their mutual coherences that contribute to the TPA process. This method is in contrast with the conventional second-order perturbation theory that uses state amplitudes and includes transition linewidths only by averaging over the final density of states. (Note that averaging over a density of states is equivalent to homogeneous broadening in lowest-order perturbation theory where there is no saturation of populations.) The conventional theory corresponds to the double quantum coherence (DQC) pathway (double-side Feynman diagram), while the additional pathways included in the density-matrix approach are the nonrephasing (NRP) and rephasing (RP) pathways.

Our treatment clarifies to what extent the predictions of these two approaches differ. We find that if the exciting field is far from resonance with any intermediate states the conventional DQC pathway typically dominates the TPA process, although under some conditions the NRP and RP pathways can make significant contributions.

The treatment we developed confirms that the DQC contribution can be greatly enhanced by the presence of frequency anticorrelations in the exciting field composed of photon pairs created in, for example, spontaneous parametric down conversion. The enhancement occurs because the frequencies of the two photons sum to that of the SPDC pump laser, so if that laser has a narrow bandwidth, the sum-frequency variable is 'compressed' into the TPA transition line profile. An equivalent explanation can be given in the time domain: the entanglement time of the photon pairs (inverse bandwidth) can be much shorter than the overall duration of the illumination pulse, meaning that the pair behaves as if confined to an ultrashort pulse whose arrive time is indeterminant. Because TPA is a nonlinear-optical, two-photon process, it is enhanced when the exciting light is confined to shorter time intervals. Detailed discussion and quantitative estimates of such effects are given in [20], and a general derivation of a hard upper bound of such enhancement is given in [21].

The enhancement of TPA probability by EPP relative to that using a coherent state with a similar pulse shapes can be quantified by, for example, combining results from Eqs.146 and 107. We see that assuming an exponential pulse shape, the DQC contribution driven by EPP is enhanced relative to the coherent-state case by the ratio, which we call the *quantum enhancement factor* [20, 21]

$$QEF^{DQC} \simeq \frac{2N_{EPP}}{N_{coh}^2} \times \frac{\sigma_B}{\Gamma} \ . \tag{161}$$

where we used $N_{EPP} = 2\varepsilon^2$. Recall that $\sigma_B$ is roughly the bandwidth of the EPP field, while $\Gamma$ is roughly the inverse of the pulse duration for both the EPP and coherent-state pulses. Similar results are found for various pulse shapes, and a general result independent of pulse shape is given in [21]. Recall the EPP results are valid under the condition of isolated EPP, that is, not overlapping photon pairs, and are therefore restricted to very low flux.

When the mean number of EPP photons in an excitation pulse is equal to the mean number of photons in a weak coherent-state pulse, $N_{EPP} = 2\varepsilon^2 = N_{coh} \equiv N_{both}$, this ratio equals $(2/N_{both}) \times (\sigma_B/\Gamma)$. Thus, if the mean number of photons in a pulse is much less than one, the first factor, which we call the 'photon-number enhancement factor,' can be large. The second factor, which we call the 'spectral enhancement factor,' can also be large if the EPP field's bandwidth is much greater than its inverse duration. This condition corresponds to large time-frequency entanglement.

Unfortunately, in many applications of interest, such as spectroscopy or two-photon microscopy, the predictions here indicate that in most practical cases the predicted final-state population is too small to be detected for typical molecules using typical technology in current use. As discussed in **Sec. 7.3**, conventional two-photon cross sections are extremely small typically. The amount of enhancement that can be achieved by the number- and frequency correlations calculated here is not likely great enough to overcome the small cross section in typical scenarios.

A major question that is addressed by the theory is to what extent the NRP and RP contributions are similarly enhanced by frequency correlations in the exciting field. We find, not surprisingly, that there is no such enhancement because these processes are step-wise, occurring through populations or mutual coherences among intermediate states. The step-wise nature of these processes disrupts the delicate frequency anticorrelations in the exciting field leading to no enhancement. The present paper is the first, to our knowledge, pointing out that the NRP and RP pathways do not provide an explanation for the anomalously large ETPA probabilities reported in some experimental studies.

This conclusion helps address the presently controversial issue discussed above concerning the detectability of entangled two-photon absorption (ETPA). Some experimental studies [15, 16] have reported apparent values of ETPA excitation probabilities that greatly exceed the values predicted here (for example, Eq.142). Other studies have recently found upper bounds on EPTA probabilities that are much smaller and in line with the present predictions [17, 18, 19], but the final word has yet to be spoken on this question.

In the following we comment on limitations and possible extensions of the theory.

First, we note that the fourth-order perturbation solution for the density matrix that is used here and in many treatments of ultrafast-laser spectroscopy is suited for excitation by short laser pulses but not applicable to calculate steady-state responses. This theory treats the final state as merely an 'integrating receptacle' for population. The challenge is to develop a non-perturbative treatment that allows the final state population to be a dynamical variable. This would allow steady-state solutions including quantum states of the exciting field, and would enable direct comparisons with semiclassical treatments such as presented in [50, 51, 52, 53, 54].

Another limitation of the present theory is that it treats only isolated photon pairs incident on the atom or molecule. Some experiments have been caried out with large fluxes such that pairs do overlap, [18, 14] and a couple of theoretical treatments cover such cases. [23, 6] The study of TPA using multi-spectral-mode squeezed states of light would be worthwhile. And, careful consideration of the roles of NRP and RP pathways in this context would be of interest.

A challenging question is to what extent the frequency correlations that enhance ETPA could be mimicked by 'classical' fluctuating fields. Frequency correlations could be built into such a model using a statistical mixture of coherent states, as in [57], but Schlawin and Buchleitner argue that such states do not enhance the absorption probability above the pure coherent-state case, [24], and Lerch and Stefanov show that a statistical mixture of correlated monochromatic states can mimic the frequency correlations but not the time correlations. [58] Furthermore, the enhancement by photon number correlations (the correlated arrival of pairs) likely cannot be mimicked perfectly by classical fields.

The treatment of collisional dephasing used here models homogeneous line shapes as Lorentzian at all frequencies. While this approach is standard and common, it is an oversimplification of the physics. There are two well-known ways to improve the treatment—the Brownian-oscillator-bath model, which is appropriate for spectroscopy of solvated molecules, [59] and the non-impact theory of collisional line broadening, which is well developed for spectroscopy of atomic or molecular vapors. [60, 61] For example, the effectiveness of collisional dephasing can be greatly reduced when light is detuned far from line center. To invert the argument, observing excitation of states far from resonance can be used to characterize the dephasing bath itself, and yield important information.

Finally, we mention the possibility to use optical phase modulation techniques similar to those used in multidimensional spectroscopy with ultrafast lasers to dissect the separate contributions of the DQC, NRP, and RP pathways. [62, 63, 64, 65, 66] If the challenge of low signal levels can be overcome, then combining EPP excitation with such phase modulation techniques might provide a new avenue for obtaining hard-to-get information on molecular structure and dynamics.

## Acknowledgements

We thank Perry Rice, Paul Berman, Jeff Cina, Daniel Steck, Steven van Enk, Markus Allgaier, Brian Smith and Sofiane Merkouche for helpful discussions. This work was supported by grants from the National Science Foundation RAISE-TAQS Program (PHY-1839216).

## Data Availability

The data supporting this study are contained within the article.

## Disclaimer

The authors declare no conflicts of interest.

================================================================

## Appendix A – Molecular dephasing and correlation functions

The molecular correlation functions introduced in Eq.(73) are evaluated by first writing them in a common form:

$$\begin{aligned} C_M(t_a, t_b, t_c, t_d) &= Tr_M\left(\hat{\mu}^{(-)}(t_a)\hat{\mu}^{(-)}(t_b)\big|g\big\rangle\big\langle g\big|\hat{\mu}^{(+)}(t_c)\hat{\mu}^{(+)}(t_d)\big|f\big\rangle\big\langle f\big|\right) \\ &= \big\langle f\big|\hat{\mu}^{(-)}(t_a)\hat{\mu}^{(-)}(t_b)\big|g\big\rangle\big\langle g\big|\hat{\mu}^{(+)}(t_c)\hat{\mu}^{(+)}(t_d)\big|f\big\rangle \end{aligned} \quad , \tag{162}$$

where:

$$\begin{aligned} \big\langle f\big|\hat{\mu}^{(-)}(t_a)\hat{\mu}^{(-)}(t_b)\big|g\big\rangle &= \sum_e \mu_{fe}e^{i(\omega_f - \omega_e)t_a}\mu_{eg}e^{i(\omega_e - \omega_g)t_b} \\ \big\langle g\big|\hat{\mu}^{(+)}(t_c)\hat{\mu}^{(+)}(t_d)\big|f\big\rangle &= \sum_e \mu_{ge}e^{i(\omega_g - \omega_e)t_c}\mu_{ef}e^{i(\omega_e - \omega_f)t_d} \end{aligned} \quad , \tag{163}$$

which gives in the three cases:

$$C_M^i(t_a, t_b, t_c, t_d) = \sum_{e,e'}\mu_{fe}\mu_{e'g}\mu_{ge}\mu_{ef}\, G^i \quad (i = DQC, NRP, RP) \quad , \tag{164}$$

where:

$$G^{DQC} = e^{i(\omega_f - \omega_e)t_4} e^{i(\omega_e - \omega_g)t_3} e^{i(\omega_g - \omega_e)t_1} e^{i(\omega_e - \omega_f)t_2}$$

$$G^{NRP} = e^{i(\omega_f - \omega_e)t_4} e^{i(\omega_e - \omega_g)t_2} e^{i(\omega_g - \omega_e)t_1} e^{i(\omega_e - \omega_f)t_3} \quad . \tag{165}$$

$$G^{RP} = e^{i(\omega_f - \omega_e)t_3} e^{i(\omega_e - \omega_g)t_2} e^{i(\omega_g - \omega_e)t_1} e^{i(\omega_e - \omega_f)t_4}$$

To apply Kubo dephasing theory (**Sec. 5.2**), we need to group the factors so we can identify disjoint time intervals in which the dephasing interactions occur. This grouping allows treating the dephasing interactions in one interval as statistically independent of those in other intervals. Thus, we introduce the difference-time variables $r = t_4 - t_3$, $s = t_3 - t_2$, $\tau = t_2 - t_1$, which means $t_1 = (t_4 - r - s - \tau)$, $t_2 = (t_4 - r - s)$, $t_3 = (t_4 - r)$. The integration ranges are $[0, \infty]$ for $r$, $s$, and $\tau$, and $[-\infty, \infty]$ for $t_4$. Then, denoting $\omega_{ij} = \omega_i - \omega_j$, we have:

$$G^{DQC} = e^{-i\omega_{ef}r} e^{-i\omega_{gf}s} e^{-i\omega_{ge}\tau}$$

$$G^{NRP} = e^{-i\omega_{ef}r} e^{-i\omega_{e'e}s} e^{-i\omega_{ge}\tau} \tag{166}$$

$$G^{RP} = e^{-i\omega_{fe}r} e^{-i\omega_{e'e}s} e^{-i\omega_{ge}\tau} \quad .$$

Note $t_4$ has dropped out in these terms. Since the time intervals are disjoint, we can apply Kubo theory in each separately, and replace:

$$G^{DQC} \rightarrow e^{-(\gamma_{fe} - i\omega_{fe})r} e^{-(\gamma_{fg} - i\omega_{fg})s} e^{-(\gamma_{eg} - i\omega_{eg})\tau}$$

$$G^{NRP} \rightarrow e^{-(\gamma_{fe} - i\omega_{fe})r} e^{-(\gamma_{ee'} - i\omega_{ee'})s} e^{-(\gamma_{eg} - i\omega_{eg})\tau} \tag{167}$$

$$G^{RP} \rightarrow e^{-(\gamma_{fe} + i\omega_{fe})r} e^{-(\gamma_{ee'} - i\omega_{ee'})s} e^{-(\gamma_{eg} - i\omega_{eg})\tau} \quad .$$

Then, from Eq.(164):

$$C_M^{DQC} = \sum_{e,e'} \mu_{fe'} \mu_{e'g} \mu_{ge} \mu_{ef} \, e^{-(\gamma_{fe} - i\omega_{fe})r} \, e^{-(\gamma_{fg} - i\omega_{fg})s} \, e^{-(\gamma_{eg} - i\omega_{eg})\tau}$$

$$C_M^{NRP} = \sum_{e,e'} \mu_{fe'} \mu_{e'g} \mu_{ge} \mu_{ef} \, e^{-(\gamma_{fe} - i\omega_{fe})r} \, e^{-(\gamma_{ee'} - i\omega_{ee'})s} \, e^{-(\gamma_{eg} - i\omega_{eg})\tau} \tag{168}$$

$$C_M^{RP} = \sum_{e,e'} \mu_{fe'} \mu_{e'g} \mu_{ge} \mu_{ef} \, e^{-(\gamma_{fg} + i\omega_{fe})r} \, e^{-(\gamma_{ee'} - i\omega_{ee'})s} \, e^{-(\gamma_{eg} - i\omega_{eg})\tau} \quad .$$

which reproduces Eq.(74).

## Appendix B – Under what conditions is the dipole matrix element product real?

In cases where the dipole matrix element product $M_{fe'eg} = \mu_{fe'} \mu_{e'g} \mu_{ge} \mu_{ef}$ is real, the analysis of expressions like those in Eq.(87) is simplified. Such a simplification is especially useful for the NRP and RP terms, where $M_{fe'eg}$ does not separate from the frequency-dependent functions as it does for the DQC case, as shown in Eq.(100). While the question is $M_{fe'eg}$ real is not easily answered in general, it is useful to consider a few special cases.

In one-dimensional systems all eigenfunctions are real, and thus $M_{fe'eg}$ is real.

For TPA in single-electron atoms, the only complex variations in the eigenfunctions enter in the form $\exp(-im\phi)$, and these factors when integrated will always produce real matrix elements.

In molecules with a high degree of symmetry, $M_{fe'eg}$ can be proven to be real, at least in special cases. Consider, for example, a symmetric $N$-mer, that is $N$ identical atoms or molecules coming together to form a symmetric structure. Denote the collective ground state, with all monomers in their lowest-energy state, by $|g\rangle$. Denote the set of singly-excited states, with one monomer in its first excited state, by $|u_n\rangle$ ($n = 1$ to $N$). And denote a particular doubly-excited state, with two monomers in their first excited state, by $|f\rangle$. Assume by symmetry that all dipole matrix elements connected to the ground state are equal: $\langle u_n|\mu|g\rangle = \mu_g$. And assume that all dipole matrix elements connected to the doubly-excited state are equal: $\langle u_n|\mu|f\rangle = \mu_f$. If the degeneracy of the singly-excited states is lifted by a symmetric interaction among the systems, the singly-excited states are mixed by a unitary transformation to create new eigenstates in the singly-excited manifold:

$$|j\rangle = \sum_{n=1}^{N} U_{jn}|u_n\rangle \ . \tag{169}$$

The dipole matrix elements transformed to:

$$\mu_{jg} = \mu_g \sum_{n=1}^{N} U_{jn}^{\ *} \ , \ \mu_{jf} = \mu_f \sum_{n=1}^{N} U_{jn}^{\ *} \ . \tag{170}$$

Then:

$$\begin{aligned} M_{fe'eg} &= \mu_{fe'}\mu_{e'g}\mu_{ge}\mu_{ef} \\ &= \left|\mu_g\right|^2 \left|\mu_f\right|^2 \sum_{m,n,r,s} U_{e'm}U_{e'n}^{\ *} \ U_{er}U_{es}^{\ *} \ . \end{aligned} \tag{171}$$

Conjugating this expression and swapping indices $n \leftrightarrow m, r \leftrightarrow s$ shows $M_{fe'eg}^{*} = M_{fe'eg}$, which verifies it is real.

We are not aware of more general proofs on the reality of $M_{fe'eg}$, but such would be useful for predicting TPA signals in complex molecules.

### Appendix C – TPA by a coherent pulse in the impulsive limit

In an extreme limit, if the coherent state is a pulse much shorter than the molecular dephasing time—the impulsive limit—then Eq.(100) becomes:

$$P_f^{DQC} = \Sigma^{(2)} L_0{}^4 \left| \int A^2(t)\, dt \right|^2 , \tag{172}$$

where we used $K(0) = \int A^2(t)\, dt$. If $A(t)$ is real and positive, then $K(0) = 1$, reproducing the result in Eq.(118) for the Gaussian coherent state in the impulsive limit $\sigma \gg \gamma_{fg}$. In this limit the probability does not depend on $\gamma_{fg}$ or $\sigma$ because for the nonlinear TPA process the effect of spreading the spectrum over a broader range is compensated by the increasing intensity in the time domain.

Note that $A^2(t)$ is proportional to the two-photon Rabi frequency [67] and can be complex. This means that the TPA probability can go to zero in the impulsive limit if the pulse constitutes a two-photon zero-$\pi$ pulse, defined by $K(0) = 0$.